\documentclass{JHEP3}
\input epsf
\usepackage{epsfig}
\usepackage{amssymb}
\usepackage[active]{srcltx}

\def\beq{\begin{equation}}
\def\be{\begin{equation}}
\def\eeq{\end{equation}}
\def\ee{\end{equation}}

\def\la{\langle}

\def\Del3{\hat{\Delta}_3}

\newcommand{\bea}{\begin{eqnarray}}
\newcommand{\eea}{\end{eqnarray}}
\newcommand{\bean}{\begin{eqnarray*}}
\newcommand{\eean}{\end{eqnarray*}}

\def\O #1{\overline{#1}}

\def\W #1{\widetilde{#1}}

\def\braket#1{\left\langle #1 \right\rangle}

\def\ket#1{\left| #1\right\rangle}
\def\gb #1{ \left\langle #1 \right]}

\def\det{\mathop{\rm det}}

\def\a{{\alpha}}

\def\la{\lambda}
\def\eps{\epsilon}

\def\vev{\braket}

\def\Del3{\hat{\Delta}_3}

\def\Label#1{\label{#1}%
  \smash{\hbox to0pt{\raise1ex\hbox{\tiny[#1]}\hss}}}

\preprint{
\\
{\tt CERN-PH-TH/2006-272}}
\title{Unitarity cuts and reduction to master integrals 
in $d$ dimensions for one-loop amplitudes}
\author{Charalampos Anastasiou\footnote{CERN, 1211 Geneva, 23, Switzerland}\,,
 Ruth Britto\footnote{Institute for Theoretical Physics, University of Amsterdam,
Valckenierstraat 65, 1018 XE Amsterdam, The Netherlands}\,,
 Bo Feng\footnote{Blackett Laboratory \& The Institute for Mathematical Sciences, Imperial College,
London, SW7 2AZ, UK\,
 and Center of Mathematical Science, Zhejiang University, Hangzhou, China}\,,
 Zoltan Kunszt\footnote{
Institute of Theoretical Physics, ETH Zurich, 8093 Zurich, Switzerland
}\, and
Pierpaolo Mastrolia\footnote{
Institute for Theoretical Physics, University of Zurich,
8057 Zurich, Switzerland}
\\
}

\abstract{We present an alternative 
reduction to master integrals 
for one-loop amplitudes using a unitarity cut method in
arbitrary dimensions. We carry out the reduction in
two steps. The first step is a pure four-dimensional cut-integration 
of tree amplitudes with a mass parameter,
and the second step is applying dimensional shift identities to master
integrals. This reduction is performed at the  integrand level, so that
coefficients can be read out algebraically.}

\begin{document}

\section{Introduction}

At the Large Hadron Collider (LHC),  complex final states will
be observed frequently and  therefore  the predictions of the Standard
Model for  the production rate of  such  events will have to be
evaluated  precisely.  This requires  the calculation
of next-to-leading order QCD corrections to the cross-section of a
large number of production processes, such as   multi-jets, vector
bosons and jets, top quarks  and  jets, etc.
We need  multileg amplitudes  at one loop accuracy.
 In the traditional method, one generates the amplitudes according to
Feynman diagrams, expresses the tensor loop integral in terms of scalar
form factors and  reduces the required scalar integrals to master integrals
using recurrence identities.
In the case of five or more particle  processes the use of this  
straightforward method becomes cumbersome.
The number of Feynman diagrams and   terms  generated have factorial 
growth. Furthermore, the presence of Gram determinants in the denominators 
of the reduction coefficients requires suitably optimized numerical 
techniques.

In the last few years remarkable progress has been achieved in replacing the traditional
method with more efficient approaches which build on the properties of
 Yang-Mills theories more explicitly. In particular, very efficient recursive algorithms
are available for the calculation of  tree amplitudes
 and  facilitate the calculation  of one loop amplitudes.\footnote{For
   reviews of this progress, see for example \cite{Cachazo:2005ga,Dixon:2005cf}.}
Many of these new ideas have been stimulated by the suggestion of Witten  to transform
QCD amplitudes to twistor space \cite{Witten:2003nn}.

The new techniques could also be extended  to the calculation of
loop amplitudes.  One can apply the unitarity cut
method~\cite{Bern:1994zx}, which uses tree amplitudes as input and so
avoids the generation of Feynman diagrams.  Recently,  the
four-dimensional unitarity cut method  has been further developed as
an  efficient systematic tool to calculate QCD
amplitudes~\cite{Britto:2004nc,Britto:2005ha}, building on techniques
inspired by twistor space
geometry~\cite{Witten:2003nn,Cachazo:2004kj,Cachazo:2004by,Brandhuber:2004yw,Cachazo:2004zb,Bena:2004xu,Cachazo:2004dr,Britto:2004nj}.
The phase-space integration  and the reduction to master integrals are
carried out explicitly in terms of spinors.  Many, mostly supersymmetric, amplitudes can be reconstructed fully with this
technique~\cite{Britto:2004nc,Britto:2005ha,Britto:2006sj}. However,
the mapping to master integrals is in general incomplete, since it
misses rational contributions that arise from multiplying $1/\eps$
poles  of master integrals with ${{\cal O}}(\epsilon)$ coefficients.  This gap has been filled recently by methods that target these
rational contributions specifically, either by developing~\cite{Bern:2005hs,Bern:2005ji,Bern:2005cq,Berger:2006ci,Berger:2006vq}
recursion relations for amplitudes~\cite{Britto:2004ap,Britto:2005fq},
or by using specialized diagrammatic
reductions~\cite{Xiao:2006vr,Su:2006vs,Xiao:2006vt,Binoth:2006hk}. As
a result, for example, short analytic formulas are now available for all the one-loop six gluon QCD helicity amplitudes.

We can, however,  reconstruct the full amplitude with the 
unitarity cut method, provided the cut integrals are treated
in $d=4-2\eps$ dimensions~\cite{vanNeerven:1985xr}.
A complete method for one-loop calculations
was developed in the pioneering work of Bern
{\em et al.}~\cite{Bern:1995db,Bern:1996je,Bern:1996ja}, and it was recently
re-examined in~\cite{Brandhuber:2005jw}. The idea is that the
$(-2\eps)$-dimensional component of momentum can be considered as
constant, and orthogonal to the $4$-dimensional components.  From this
point of view, a massless $d$-dimensional scalar can be traded for a
massive $4$-dimensional scalar.  Then, unitarity cuts can be applied
to constrain the coefficients of master integrals.    
However, the calculation  of unitarity cuts
is generally difficult because of the reduction
of $d$-dimensional tensor integrals. Eventually, in the papers
mentioned above  one resorts to traditional  reduction
methods to complete the computation.

In~\cite{Anastasiou:2006jv} we have reported  results on  an efficient implementation
of the  $d$-dimensional unitary cut
method.   One-loop  amplitudes can be reduced  to master integrals
for arbitrary values of the dimension parameter. An important result is that
we can read out the coefficients of the master integrals without fully
carrying out the $d$-dimensional phase space integrals.
The problem is reduced to four dimensional integration, where we can
now capitalize on the recent advances in computation.
The four dimensional tensor integrals can be calculated
 using spinor integration for
light-like
momenta~\cite{Cachazo:2004dr,Britto:2004nj,Britto:2004nc,Britto:2005ha,Britto:2006fc}.
Recursion relations for amplitudes with massive scalars have been developed
\cite{Badger:2005zh} specifically for generating the tree-level input
for $d$-dimensional unitarity cuts.

In this paper we outline the $d$-dimensional unitarity cut method in
detail and give simple illustrative applications using spinorial integration
for tensor reduction.  Here we work entirely in terms of standard
double cuts; some work with generalized unitarity cuts in $d$
dimensions has appeared in
\cite{Brandhuber:2005jw,Mastrolia:2006ki,Britto:2006fc}.  A different
method of constructing the master integral coefficients of one-loop
amplitudes, from the values of the loop momentum that correspond to
unitarity cuts, has been presented in \cite{Ossola:2006us}.

In section 2 we discuss the parametrization of $d$-dimensional cut
integrals.  We identify the 4-dimensional integral within the
$d$-dimensional integral, leaving the final $(-2\eps)$-dimensional integral
for last.  The 4-dimensional integral can be performed by a method of
choice; here we proceed 
in terms of spinorial variables.  We show the cut bubble as a
prototype of any cut integral and then set up the integral for a
general amplitude. 

In section 3, we derive the
integral representations of the cut master integrals, namely scalar
bubbles, triangles, boxes, and pentagons.   The physical arguments of
\cite{Britto:2005ha,Britto:2006sj,Britto:2006fc} state that all
possible integrands are related to these basis integrals simply by
polynomial factors in the  $(-2\eps)$-dimensional mass parameter.  We
relate these general integrands to the basis integrands by dimensional
shift identities, which here take the form of recursion and reduction
relations.  (Recursion refers to the degree of the polynomial.)
We derive these identities and explain their application.

In section 4, we work through the examples of the five-gluon all-plus
amplitude and the four-gluon amplitudes and confirm that our results
agree with \cite{Bern:1995db,Bern:1996ja,Brandhuber:2005jw}.
The reduction is done using spinorial integration
\cite{Britto:2005ha,Britto:2006sj}.  These spinorial integrals are evaluated using Schouten identities,
Feynman parameter integrals  and the holomorphic anomaly formula \cite{Cachazo:2004by}.
Since in the $d$-dimensional unitarity  method the integrand of spinorial integrals depends
on an additional mass parameter, the size of the expressions is
larger and the recognition of the
scalar master integrals is more involved than in the four-dimensional
case.

Appendix A discusses the kinematic region and domains of integration
for a unitarity cut.  Appendix B gives further details of the various
master integrals.  Appendix C contains helpful identities and
formalisms for spinor integration, in particular with regard to
quadratic denominator factors.

\section{The $d$-dimensional unitarity method}

The $n$-point scalar function is defined by\footnote{ Our convention
  is to omit the prefactor $i(-1)^{n+1}(4\pi)^{D/2}$ that is common
  elsewhere in the literature.}
\bea I_n & = & \int {d^{4-2\eps} p \over (2\pi)^{4-2\eps}}{1\over
p^2 (p-K_1)^2 (p-K_1-K_2)^2... (p-\sum_{j=1}^{n-1}
K_j)^2}.~~~\label{n-scalar} \eea
We operate in the four-dimensional helicity (FDH) scheme, in which all external momenta are in four dimensions.  In this formula, therefore, the loop momentum $p$ is $(4-2\eps)$-dimensional, while all the
$K_i$ are $4$-dimensional. We can decompose $p=\W \ell+\vec{\mu}$
where $\W \ell$ is $4$-dimensional and $\vec{\mu}$ is
$(-2\eps)$-dimensional. Then
the integration measure becomes
\bean \int {d^{4-2\eps} p\over (2\pi)^{4-2\eps}} & = & \int {d^{4}
\W\ell\over (2\pi)^{4}}\int {d^{-2\eps} \ell_\eps\over
(2\pi)^{-2\eps}} = \int {d^{4} \W\ell\over (2\pi)^4}
{(4\pi)^{\eps}\over \Gamma(-\eps)} \int  d\mu^2 ~
(\mu^2)^{-1-\eps},\eean
and the scalar function is
\bea I_n & = & {(4\pi)^{\eps}\over \Gamma(-\eps)} \int  d\mu^2
(\mu^2)^{-1-\eps} \int {d^{4} \W\ell\over (2\pi)^4}{1\over (\W
\ell^2-\mu^2) ((\W \ell-K_1)^2-\mu^2)... ((\W \ell-\sum_{j=1}^{n-1}
K_j)^2-\mu^2)}.~~~\label{n-scalar-1} \eea
We will use spinor integration when we cut the
4D momentum $\W \ell$, so we choose to decompose it into a linear combination of a light-like momentum variable and a fixed vector $K$:
\bea \W \ell= \ell+z K,~~~~~\ell^2=0,~~~\Longrightarrow \int d^4\W
\ell= \int dz ~d^4\ell ~\delta^+(\ell^2) (2 \ell \cdot K).
~~~\label{changing} \eea
Eventually it will be convenient to choose $K$ to be the momentum through the unitarity cut.
This is one of the most important ideas that enables our whole program to work.
Further we define
\bea u= {4\mu^2\over K^2}.~~~~\label{def-u}\eea
As will be clear from  discussions in Appendix A, in our cut
calculation we have $u\in [0,1]$. Therefore
\bea  {(4\pi)^{\eps}\over (2\pi)^4\Gamma(-\eps)} \int  d\mu^2 ~
(\mu^2)^{-1-\eps}\to {(4\pi)^{\eps}\over (2\pi)^4\Gamma(-\eps)}\left(
{K^2\over 4}\right)^{-\eps} \int_0^1 du~
u^{-1-\eps}.~~~\label{change-to-u}\eea
Since ${(4\pi)^{\eps}\over (2 \pi)^4\Gamma(-\eps)}\left( {K^2\over
4}\right)^{-\eps}$ is an universal factor
appearing on both sides of every cut calculation,
 we will neglect it throughout the rest of the paper.

Finally we arrive at the equation
\bea I_n & = & \int_0^1 du~ u^{-1-\eps} \int dz ~ d^4\ell
 ~ \delta^+(\ell^2) (2 \ell \cdot K) {1\over (\W \ell^2-\mu^2) ((\W
\ell-K_1)^2-\mu^2)... ((\W \ell-\sum_{j=1}^{n-1}
K_j)^2-\mu^2)},~~~\label{n-scalar-2}\eea
where $\mu^2$ is related to $u$ through (\ref{def-u}).

At this point we are ready to carry out the 4-dimensional cut integration.  We do this in the language of spinors.

\subsection{The cut-integration of bubble functions}

The cut of a scalar bubble is the simplest kind of unitarity cut, so it is instructive as well as useful to go through this case in detail.  Then we will be able to set up the framework for any other cut of master integrals or amplitudes.

The expression of the (double) cut of the bubble function is given by (\ref{n-scalar-2})
\bea C[I_2(K)] & = & \int_0^1 du~ u^{-1-\eps} \int dz~ d^4\ell ~
\delta^+(\ell^2) (2 \ell \cdot K) \delta(\W \ell^2-\mu^2) \delta((\W
\ell-K)^2-\mu^2),
\eea
where $\W \ell=\ell+z K$ with $\ell^2=0$, and $\mu^2$ is related to
$u$ by (\ref{def-u}).
We make use of the delta functions to rewrite the integral as follows.
\bean C[I_2(K)] & = &
\int_0^1 du~ u^{-1-\eps} \int dz~ d^4\ell
~\delta^+(\ell^2) (2 \ell \cdot K) \delta(\W \ell^2-\mu^2) \delta(
K^2-2K\cdot \W\ell) \\
& = & \int_0^1 du~ u^{-1-\eps} \int dz~
d^4\ell ~\delta^+(\ell^2) (2 \ell \cdot K) \delta(z^2 K^2+2z K\cdot
\ell-\mu^2) \delta( (1-2z) K^2-2K\cdot \ell)\\
& = & \int_0^1 du~
u^{-1-\eps} \int dz  (1-2z) K^2\delta(z(1-z)K^2-\mu^2) \int d^4\ell
~\delta^+(\ell^2)\delta( (1-2z) K^2-2K\cdot \ell).
\eean
Here we have brought the integral into a form where one of the
delta-functions, $\delta(z(1-z)K^2-\mu^2)$, does not depend on $\ell$.
Now we continue by transforming the integral to spinor coordinates \cite{Cachazo:2004kj}:
\bea
\ell = t\lambda\W\la,
\eea
so that the measure transforms as
\bea
\int d^4\ell \delta^{(+)}(\ell^2) ~ (\bullet ) =
\int_0^{\infty}dt~t\int\vev{\lambda~
d\lambda}[\tilde\lambda~d\tilde\lambda] ( \bullet ).
\eea
Here $t$ ranges over the positive real line, and $\la,\W\la$ are
homogeneous spinors, also written respectively as $|\ell\rangle,
|\ell]$ in many expressions involving spinor products.  The first step
in spinor integration is to integrate over the variable $t$.  This is
never true integration, because all we need is to solve the delta
function of the {\em second} cut propagator.
Thus we find
\bean C[I_2(K)] & = & \int_0^1 du~ u^{-1-\eps} \int dz (1-2z)
K^2\delta(z(1-z)K^2-\mu^2) \\ & & ~~\int \braket{\ell~d\ell}[\ell~d\ell] \int
dt~ t ~\delta( (1-2z) K^2-2K\cdot \ell).
\eean
The spinor and $t$-integrations are
similar to the four-dimensional case \cite{Britto:2005ha,Britto:2006sj}; the only new feature is the factor of
 $(1-2z)$. After this integration, we get\footnote{In this
paper we take residue instead of the negative of residue when we do
phase space integration. This is just a matter of convention because when we
calculate both sides of the cut equation, the sign cancels out.}
\bean C[I_2(K)] & = & \int_0^1 du~ u^{-1-\eps} \int dz (1-2z)
K^2\delta(z(1-z)K^2-u {K^2\over 4}) (1-2z).
\eean
Using the formula
\bea \delta(g(x))= \sum_i {\delta(x-x_i)\over
|g'(x_i)|},
~~~~~\label{delta-int} \eea
where the $x_i$'s are the roots of $g(x)$, we can finish the
$z$-integration to get
\bea C[I_2(K)] & = & \int_0^1 du~ u^{-1-\eps}
\sqrt{1-u},~~~~\label{I2-cut-1}\eea
where we have used the fact (see Appendix A) that only one
root is allowed, specifically, $z=(1-\sqrt{1-u})/2$. Equation
(\ref{I2-cut-1}) is simple enough that we can  finish the $u$
integration directly to find\footnote{The following
analytic expression is right only when ${\rm Re}(\eps)<0$. This is
the condition  for us to use  integration by parts to derive all
recursion and reduction formulas.}
\bea C[I_2(K)] & = & {\sqrt{\pi} \Gamma(-\eps) \over 2
\Gamma({3\over 2}-\eps)},~~~~{\rm
Re}(\eps)<0.~~~~\label{I2-cut-2}\eea
We can check the result (\ref{I2-cut-2})  on the well known scalar
bubble function given by
\bea I_2^{old}(K^2) & = & {r_\Gamma \over \eps (1-2 \eps)}
\left(-K^2 \right)^{-\eps}, ~~~ r_\Gamma= {\Gamma(1+\eps) \Gamma^2
(1- \eps) \over \Gamma(1-2 \eps)}. ~~~~\label{I2bubble}\eea
To take the imaginary part we need to use\footnote{To compute the
cut with momentum $K$, we work in the kinematic region where only
$K^2>0$  and all other momentum invariants are negative.}
 \bea {\rm Im}
(-K^2)^{-\eps} = 2i \sin(\pi \eps) |K^2|^{-\eps},
~~~\label{Imaginary}\eea
thus
 \bea C[I_2^{old}(K^2)] ={2i \sin(\pi \eps)\Gamma(1+\eps) \Gamma^2
(1- \eps) \over \eps (1-2 \eps)\Gamma(1-2 \eps)}
(K^2)^{-\eps}.
~~\label{nor-4}\eea
When we try to compare with our new result  (\ref{I2-cut-2}), we
must multiply (\ref{I2-cut-2}) by the following two factors: (1)
${(4\pi)^{\eps}\over \Gamma(-\eps)}\left( {K^2\over
4}\right)^{-\eps}$ from the discussion below (\ref{change-to-u});
(2) $i (4\pi)^{2-\eps}$ from our non-standard definition of the scalar
function in (\ref{n-scalar}). Considering these two facts, one can check
that ${ C[I_2^{our}(K^2)]\over
C[I_2^{old}(K^2)]}=8\pi$, which is just a matter of a different normalization when
we take $\int d^4 \ell ~\delta^+(\ell^2)$.

\subsection{Cut-integral of an amplitude}

Now we discuss how to apply the integration technique to the cut of an
amplitude.
 The general expression will be
\bea C & = & \int_0^1 du~ u^{-1-\eps} \int dz~ (1-2z)
K^2\delta(z(1-z)K^2-\mu^2) \nonumber \\ & & \int d^4 \ell ~
\delta^+(\ell^2)\delta( (1-2z) K^2-2K\cdot \ell) A_L (\W \ell_1, \W
\ell_2)A_R (\W \ell_1, \W \ell_2),
~~~\label{Gen-fra}\eea
where $A_L, A_R$ are the  tree-level amplitudes on either side of the cut.
In this
formula, $K$ is the cut-momentum and  $\W \ell_1$ and $\W \ell_2$
are the (massive) cut 4D-momenta
satisfying
\bea \W \ell_2= K-\W \ell_1,~~~\W \ell_1^2=\W
\ell_2^2=\mu^2,~~~\W\ell_1= \ell+z K.~~~~\label{gen-var}\eea

Now we explain the meaning of the expression (\ref{Gen-fra}). The
second line is simply  a 4D cut-integration  that
depends on the parameter $z$.  The techniques developed in ~\cite{Britto:2005ha,Britto:2006sj,Britto:2006fc}
 can be applied directly.
Then this result can be put into the first line, and the
$z$-integration  can be performed trivially by using the
delta-function.
 We arrive at the final expression
\bea C & = & \int_0^1 du~u^{-1-\eps} \int d^4 \ell ~
\delta^+(\ell^2)\delta( (1-2z) K^2-2K\cdot \ell) A_L (\W \ell_1, \W
\ell_2)A_R (\W \ell_1, \W \ell_2),~~~~\label{Gen-exp}\eea
where
\bea  \mu^2={K^2 \over 4}u,~~~~~~~z={1-\sqrt{1-u}\over 2}.~~~~~\label{value}\eea
This is our setup for all calculations in this paper.

\section{Identifying integrands:  cuts of master integrals}

In this section we study the cuts of the master integrals.
Our aim is to relate these with cuts of the amplitude, at the integrand level, so that we can read off the coefficients.

The general integrand arising from the cut of an amplitude looks like a series of terms that are related to cuts of master integrals by factors that are polynomial in $u$.  Therefore, we define classes of integrals related to the cuts of master integrals by additional powers of $u$.  Through integration by parts, the powers of $u$ can be stripped away.  The result is a set of ``recursion and reduction identities'' that relate any integrand to cuts of master integrals.  With these identities it is possible to read off the coefficients without any actual integration.

\subsection{Cut bubbles}

Here we consider the whole class of integrands that will be related to bubbles by
a recursion formula.
For $n \geq 0$, we define the following new
function.
\bea   {\rm Bub}^{(n)} \equiv \int_0^1 du~u^{-1-\eps} u^n
\sqrt{1-u}.
~~\label{I-2m-n}\eea
The physical cut of the bubble master integral is $C[I_2(K)]={\rm Bub}^{(0)}$.
The function ${\rm Bub}^{(n)}$
represents a term that may arise from a general cut amplitude.
It is simple enough to evaluate this integral directly, but what we want is to relate it to the master integral.  We carry this out in rather general terms, to illustrate the idea for the more complicated master integrals.
Let us see how to find a recursion relation in $n$ and eventually write  ${\rm Bub}^{(n)}$ in terms of  ${\rm Bub}^{(0)}$.
For $n\geq 1$, we integrate by parts to get
\bean {\rm Bub}^{(n)} & = &  \left. -{2\over
3}(1-u)^{3/2}u^{-1-\eps} u^n \right|_0^1 +\int_0^1 du~ {2\over 3}(n-1-\eps)(1-u)^{3/2}
u^{-1-\eps} u^{n-1} \\
& = & \int_0^1 du~ {2\over 3}(n-1-\eps)\sqrt{1-u}(1-u) u^{-1-\eps}
u^{n-1} \\
&=& {2\over 3}(n-1-\eps)
({\rm Bub}^{(n-1)}-{\rm Bub}^{(n)}).
\eean
The boundary term vanishes because ${\rm Re}(\eps)<0$.  From this we get the following recursion relation.
\bea  {\rm Bub}^{(n)}= { (n-1-\eps)\over (n+{1\over
2}-\eps)} {\rm Bub}^{(n-1)}.~~~\label{I-2m-recu-1}\eea
This recursion is easily solved.
We write the solution in the form
\bea  {\rm Bub}^{(n)} & = & F_{2\to 2}^{(n)}
 {\rm Bub}^{(0)},~~~\label{I-2m-recu-3} \eea
where the form factor is
\bea F_{2\to 2}^{(n)} & = & {\Gamma(3/2-\eps) \Gamma(n-\eps)\over
\Gamma(-\eps)\Gamma(n+3/2-\eps)}.~~~\label{F-2n-2n} \eea
Notice that this form factor does not depend on any kinematical
variables.

There is another expression for ${\rm Bub}^{(n)}$, obtained by a
different choice of integration by parts.  
\begin{equation}  {\rm Bub}^{(n)}  =  \left. {u^{n-\eps}\over n-\eps} \sqrt{1-u} \right|_0^1
+{1\over 2(n-\eps)} \int_0^1 du~{ u^{n-\eps}\over \sqrt{1-u}}
\nonumber  =  {1\over 2(n-\eps)} \int_0^1 du~{ u^{n-\eps}\over
\sqrt{1-u}}.
~~~\label{I-2m-another}\end{equation}
It is useful to be able to recognize this alternative expression when it
shows up as an
integrand.  It is the same integral found in one-mass and two-mass
triangles (for details, see Appendix B).

\subsection{Cut triangles}

We label the triangle such that the cut momentum is  $K=K_1$.  Then
 the cut-integrand is given by
\bean { \delta(\W \ell^2-\mu^2) \delta((\W \ell-K_1)^2-\mu^2)\over
((\W \ell+K_3)^2-\mu^2)}.
\eean
Using the general integration measure of 
(\ref{Gen-exp}), we get
\bean C[I_{3}(K_1;K_3)]=\int_0^1 du~ u^{-1-\eps} \int
\vev{\ell~d\ell}[\ell~d\ell]\int dt~ {t ~\delta( (1-2z)K_1^2
+t\gb{\ell|K_1|\ell}) \over K_3^2+2 z K_1\cdot K_3
-t\gb{\ell|K_3|\ell}}.
~~~\label{I-3-1}\eean
After $t$-integration we get
\bea C[I_{3}(K_1;K_3)] &=& -\int_0^1 du~ u^{-1-\eps} \sqrt{1-u}\int
\vev{\ell~d\ell}[\ell~d\ell]{1\over
\gb{\ell|K_1|\ell}\gb{\ell|P_1|\ell}} \\ &=& -\int_0^1 du~ u^{-1-\eps}
\sqrt{1-u} \int_0^1 dx {1\over P^2},
~~~\label{I-3-2}\eea
with
\bea P_1 &=& {K_3^2+2  z K_1\cdot K_3\over K_1^2} K_1+ (1-2z)
K_3, \\ P &=& x P_1-(1-x) K_1
\eea
After some algebraic manipulations we reach
\bea C[I_{3}(K_1;K_3)]=-\int_0^1 du~ u^{-1-\eps}{1\over
\sqrt{\Delta_3}}\ln \left( {Z +\sqrt{1-u}\over Z-\sqrt{1-u}
}\right),~~~\label{I-3-3}\eea
with
\bea Z=-{K_1\cdot K_3+ K_3^2 \over \sqrt{ (K_1\cdot K_3)^2-
K_1^2 K_3^2}}, ~~~~\Delta_3=4[(K_1\cdot K_3)^2- K_1^2
K_3^2].~~~\label{I3m-para}\eea
It can be shown that in our kinematic region, in which only
$K_1^2>0$ and all other momentum invariants are negative, we will have  $Z\geq 1$.

It is hard to evaluate the integral over $u$ for
(\ref{I-3-3}), but our strategy is that we {\it never need to
evaluate it.} While keeping it in integral form, we will be able to
reduce general integrands by our recursion and reduction formulas.

\vspace{0.3cm}
{\bf Recursion and reduction for triangles:}

We define the following dimensionless integrals for all nonnegative
integers $n$:
\bea  {\rm Tri}^{(n)}(Z) \equiv \int_0^1 du~u^{-1-\eps} u^n\ln \left(
{Z +\sqrt{1-u}\over Z-\sqrt{1-u} }\right).~~~\label{I-3n}\eea
 The physical cut is
\bea
C[I_3(K_1,K_3)] = -{1 \over \sqrt{\Delta_3}} {\rm Tri}^{(0)}(Z),~~\label{phys-tri}
\eea
 if we take the $Z$ and $\Delta_3$ defined in (\ref{I3m-para}). The definition
(\ref{I-3n}) was chosen because it is free of dimensional factors.
For $n\geq 1$ we can do the following integration by parts.
\bean {\rm Tri}^{(n)}(Z) & = & \left. u^{n-1-\eps} \left(
(Z^2-1+u)\ln\left( {Z +\sqrt{1-u}\over Z-\sqrt{1-u} }\right)-2
Z\sqrt{1-u}\right)\right|_0^1 \\& & -\int_0^1 du~{u^{n-2-\eps}( n-1-\eps)}
\left( (Z^2-1+u)\ln\left( {Z +\sqrt{1-u}\over Z-\sqrt{1-u}
}\right)-2 Z\sqrt{1-u}\right).
\eean
From this we derive the following recursion and reduction relation.
\bea {\rm Tri}^{(n)}(Z)= -{(Z^2-1) (n-1-\eps)\over
(n-\eps)}{\rm Tri}^{(n-1)}(Z) + {2 Z (n-1-\eps)\over
(n-\eps)}{\rm Bub}^{(n-1)}.
~~~\label{I-3n-rec-1}\eea
Here the last term in (\ref{I-3n-rec-1}) is the same  one
defined in (\ref{I-2m-n}), which is related to cut
bubbles.

The result (\ref{I-3n-rec-1}) plays two roles. First, it is the
recursion formula for coefficients of triangles. After $n$  steps in the recursion
we arrive at $n=0$, which is related to the cut of the scalar triangle by the factor $-1/\sqrt{\Delta_3}$. Second, it
establishes the reduction relation for tensor triangles to scalar
bubbles. For a given triangle, there is  only one bubble that can result from reduction, consistent with a given cut-momentum (here $K_1$).

Now we solve (\ref{I-3n-rec-1}) and get
\bea {\rm Tri}^{(n)}(Z) & = &  F_{3\to 3}^{(n)}(Z){\rm Tri}^{(0)}(Z)
+\W F_{3\to 2}^{(n)}(Z){\rm Bub}^{(0)},~~~\label{I-3n-rec-2} \eea
in terms of two form factors, which are functions of only one variable $Z$,
the identifier of a given triangle.  Explicitly, these form factors
are given by
\bea  F_{3\to 3}^{(n)}(Z) & = & {-\eps\over n-\eps} (1-Z^2)^n,
~~~\label{F-3n-3n}
\\
\W F_{3\to 2}^{(n)}(Z) & = & {1\over n-\eps} {\Gamma(3/2-\eps)\over
\Gamma(-\eps)} \sum_{k=1}^n {2 Z}(1-Z^2)^{n-k} {\Gamma(k-\eps)\over
\Gamma(k+1/2-\eps)}. ~~~\label{F-3n-2n} \eea
Equation (\ref{I-3n-rec-2}) is not
in the exact form that we want.  We need to return to the language of
physical cuts by including the factor $-1/\sqrt{\Delta_3}$
from (\ref{phys-tri}).  The recursion/reduction formula that we need
is thus:
\bea \int_0^1 du~u^{-1-\eps} u^n\left[
-{1\over\sqrt{\Delta_3}}
\ln \left( {Z +\sqrt{1-u}\over Z-\sqrt{1-u}
}\right)\right]= F_{3\to 3}^{(n)}(Z) C[I_3(K_1,K_3)]+ F_{3\to
2}^{(n)}(K_1,K_3)C[I_2(K_1)],~~~\label{I-3n-rec-3}\eea
where
\bea F_{3\to 2}^{(n)}(K_1,K_3))= -{1\over
\sqrt{\Delta_{3}}}\W F_{3\to 2}^{(n)}(Z).\eea

The relation (\ref{I-3n-rec-3}) is the main result of this subsection.
Let us comment briefly on how it will be used.  Our $d$-dimensional
unitarity cut method separates the complete cut integration
into the $u$-integral,
$\int_0^1 du u^{-1-\eps}$, and the massive 4D part.
After doing the 4D integral, we come to an expression of the form
 $\int_0^1 du u^{-1-\eps}\sum_{i\in basis}
f_i(u) C[I_i]$, where $C[I_i]$ is the cut of master integral $I_i$ and
$f_i(u)=\sum_i a_i u^i$ is the polynomial of $u$. Then we know
immediately that this term will contribute $\sum_{n} a_i F_{3\to
2}^{(i)}(K_1,K_3)$ to the bubble coefficient and $\sum_{n} a_i
F_{3\to 3}^{(i)}(Z)$ to the coefficient of the triangle.

As in the bubble case, we derive a useful identity by integrating by parts in a different way:
\bean {\rm Tri}^{(n)}(Z) & = &
\left. {u^{-1-\eps} u^{n+1}\over n-\eps}\ln \left( {Z +\sqrt{1-u}\over
Z-\sqrt{1-u} }\right)\right|_0^1-\int_0^1 du~ {u^{-1-\eps} u^{n+1}\over
n-\eps} {Z\over \sqrt{1-u} (1-u-Z^2)}.\eean
Again, since we have $Re(-\eps)>0$, the boundary contribution is zero, so
we end up with
\bea {\rm Tri}^{(n)}(Z) & = &-{Z\over
n-\eps}\int_0^1 du~ u^{-1-\eps} {u^{n+1}\over\sqrt{1-u}  (
1-u-Z^2)}.~~~~\label{I-3m-gen}\eea
%

\subsection{Cut boxes}

In this subsection we will deal with box functions.
There are several different cuts, but we would like to simplify calculations by representing them collectively  by the same expression:
\bean  {\delta(\W \ell^2-\mu^2)\delta((\W\ell-K)^2-\mu^2) \over
((\W\ell-P_1)^2-\mu^2)((\W\ell-P_2)^2-\mu^2)},
\eean
where for different cuts we need to take different values of $P_1, P_2$. To
be clear, we list the six possible cuts of a box in Table (\ref{Box-cut-table}),
with $K_1, K_2, K_3, K_4$ in clockwise ordering.  There will be two cut triangles related to each cut box consistent with the cut momentum $K$; these are indicated here as well, for use in the reduction relations.
\bea
\begin{array}{|c|c|c|c|c|} \hline
Box~Cut~K & ~~~P_1~~~ & ~~~P_2~~~ & Triangle~One's~(K_1, K_3) &
Triangle~Two's~(K_1, K_3)
\\ \hline K_1 & K_{12} & -K_4 &  (K_1, K_{34}) & (K_1, K_4) \\
\hline K_2 & K_{23} & -K_1 & (K_2, K_{41}) & (K_2, K_1) \\
\hline K_3 & K_{34} & - K_2 & (K_3, K_{12}) & (K_3, K_2) \\
\hline K_4 & K_{41} & - K_3 & (K_4, K_{23}) & (K_4, K_3) \\
\hline K_{12} & K_1  & - K_4 & (K_{34}, K_2) & (K_{12}, K_4) \\
\hline K_{23} & K_2 & - K_1 & (K_{41}, K_3) & (K_{23}, K_1) \\
\hline
\end{array}~~~\label{Box-cut-table}
\eea
After performing the $t$-integration we get
\bean C[I_{4}(K; P_1, P_2)] & = & \int_0^1 du~ u^{-1-\eps}
{(1-2z)\over K^2}\int \vev{\ell~d\ell}[\ell~d\ell] { 1\over
\gb{\ell|Q_1|\ell}\gb{\ell|Q_2|\ell}} \nonumber \\ & = & \int_0^1 du~
u^{-1-\eps} {\sqrt{1-u}\over K^2} \int_0^1 dx {1\over Q^2},
\eean
where
\bea
Q=x Q_1+(1-x) Q_2
\eea
and
\bea Q_i= -(1-2z) P_i+{ P_i^2-2z P_i\cdot K\over K^2}
K.~~\label{R1R2} \eea

For future convenience let us give the names $R_1,R_2$ to these vectors at the point $u=0$:
\bea
R_i \equiv  - P_i+{ P_i^2 \over K^2}
K.~~\label{actR1R2}
\eea
 We define some additional variables:
\bea
& \alpha_i \equiv R_i \cdot K,~~~~  & \beta_i \equiv P_i^2 - {(P_i \cdot K)^2 \over K^2}
\nonumber \\
& A \equiv -{1\over K^2}\det\left( \begin{array}{ccc} P_1^2~~ & P_1\cdot
P_2~~ & P_1\cdot K \\ P_1\cdot P_2~~ & P_2^2~~ & P_2\cdot K \\
P_1\cdot K~~ & P_2\cdot K~~ &
K^2\end{array}\right),~~~~~~~~  & C \equiv {1\over K^2}\det\left(
\begin{array}{cl}  P_1\cdot P_2~~ & P_1\cdot K \\ P_2\cdot K~~ & K^2
\end{array}\right), \nonumber \\
& B \equiv -\det \left( \begin{array}{cc}
R_1^2 & R_1\cdot R_2 \\
R_1\cdot R_2 &
R_2^2
\end{array}\right),~~~~~~~~~~~
 & D \equiv R_1\cdot R_2.
~~~~\label{Para-3}\eea

Using the identities
\bea  Q_i^2 & = & (1-u) \beta_i +{\alpha_i^2\over K^2}, \nonumber \\
Q_1\cdot Q_2 & = & (1-u) C +{ \alpha_1 \alpha_2\over K^2}~, \nonumber \\
(Q_1\cdot Q_2)^2-Q_1^2 Q_2^2 & = & (1-u) (B - A u),~~~\label{boxdisc}\eea
 we can derive
\bea \int_0^1 dx {1\over Q^2} & = & {1\over 2\sqrt{ (Q_1\cdot
Q_2)^2- Q_1^2 Q_2^2}} \ln {(Q_1\cdot Q_2) +\sqrt{ (Q_1\cdot
Q_2)^2- Q_1^2 Q_2^2}\over (Q_1\cdot Q_2)-\sqrt{ (Q_1\cdot
Q_2)^2- Q_1^2 Q_2^2}},~~~\label{R-form-1}\\& = & {1\over
2\sqrt{1-u}\sqrt{ B - A u}} \ln \left( {D - C u+ \sqrt{1-u}\sqrt{
B - A u}\over D - C u- \sqrt{1-u}\sqrt{
B - A u}}\right),~~~\label{R-form-2}\eea
where from (\ref{R-form-1}) to (\ref{R-form-2}) we have worked out
the $u$-dependence.

Finally we arrive at the expression
\bea C[I_{4}(K; P_1, P_2)] & = & {1\over 2K^2} \int_0^1 du~
u^{-1-\eps} {1\over \sqrt{ B - A u}} \ln \left( {D - C u+
\sqrt{1-u}\sqrt{ B - A u}\over D - C u-\sqrt{1-u}\sqrt{
B - A u}}\right).~~~\label{B-cut-gen}\eea

\vspace{0.3cm}
{\bf Recursion and reduction for boxes:}

We define the following dimensionless integrals for all nonnegative
integers $n$:
\bea {\rm Box}^{(n)}(A,B,C,D) & \equiv &  \int_0^1 du~ u^{-1-\eps}
{u^n\over \sqrt{ B - A u}} \ln \left( {D - C u+ \sqrt{1-u}\sqrt{
B - A u}\over D - C u- \sqrt{1-u}\sqrt{ B - A u}}\right).
~~~\label{4m-n} \eea
The physical cut is
\bea
C[I_{4}(K;P_1,P_2)] = {1 \over 2K^2}{\rm Box}^{(0)}(A,B,C,D),
\eea
 if $A,B,C,D$ are defined as in (\ref{Para-3}).

To derive recursion and reduction relations, we select the factor $1/\sqrt{ B - A u}$ and integrate by parts. The boundary term is zero if
$n\geq 1$; thus we have
\bea {\rm Box}^{(n)}(A,B,C,D) & = &{ 2 (n-1-\eps) \over A} ( B ~{\rm
Box}^{(n-1)}(A,B,C,D) -A ~{\rm Box}^{(n)}(A,B,C,D) ) +
T,~~\label{Box-rec-1}\eea
with
\bea T & = & -{2\over A } \int_0^1 du~ {u^{n-1-\eps}\over \sqrt{1-u}}
{
(A D - 2 B C + B D) - u (2 A D - A D - B C)
\over (D - C u)^2- (1-u) (B - A u)}.
~~~\label{T-1}\eea
This quantity $T$ is related to triangle integrals.  To see this, factorize the quadratic polynomial in the denominator:
\bean (D - C u)^2- (1-u) (B - A u)=
\beta_1 \beta_2
(1-u- Z_1^2) (1-u-
Z_2^2), \eean
with
\bea Z_i^2= -{
\alpha_i^2 \over \beta_i
 K^2}.
~~~\label{A1-A2} \eea
Comparing with (\ref{I3m-para})  we can see that these $Z_1^2$ and
$Z_2^2$ correspond to  the $Z^2$ of some triangle functions. Here in particular,
\bea
Z_i = -{R_i \cdot K \over \sqrt{(P_i \cdot K)^2-P_i^2K^2}}.~~~\label{derived-z}
\eea

In Table
(\ref{Box-cut-table}) we have listed which kinds of triangles a box
with given cut would reduce to.

Using the above result with the cut-triangle expression from (\ref{I-3m-gen}) we get
\bea T & = & {2\over A } \int_0^1 du~ {u^{n-1-\eps}\over
\sqrt{1-u}}\left( {C_{Z_1}  \over 1-u- Z_1^2}+{ C_{Z_2}\over 1-u-
Z_2^2}\right) \nonumber \\
& = & -{2C_{Z_1} (n-1-\eps)\over Z_1 A }
{\rm Tri}^{(n-1)}(Z_1)-{2C_{Z_2}(n-1-\eps)\over Z_2 A } {\rm Tri}^{(n-1)}(Z_2),
\eea
where $Z_1, Z_2$
should be derived from (\ref{derived-z}) with reference to Table (\ref{Box-cut-table}), and $C_{Z_1}$ and
$C_{Z_2}$ are given by
\bea C_{Z_i}= {D + (Z_i^2-1) C}.~~~\label{Coeff-A12}\eea

Putting it all together we find the  recursion and reduction formulas.
\bea {\rm Box}^{(n)}(A,B,C,D) & = &  {(n-1-\eps)\over (n-{1\over
2}-\eps)} {B\over A} {\rm Box}^{(n-1)}(A,B,C,D) + {1 \over
2(n-{1\over
2}-\eps)}T\\
&=& {(n-1-\eps)\over (n-{1\over 2}-\eps)}
{B\over A} {\rm Box}^{(n-1)}(A,B,C,D)~~\label{Box-rec-2} \\
& &
 -{(n-1-\eps)C_{Z_1}\over (n-{1\over
2}-\eps)A~ Z_1 } {\rm Tri}^{(n-1)}(Z_1)-{(n-1-\eps)C_{Z_2}\over (n-{1\over 2}-\eps)A~ Z_2
}{\rm Tri}^{(n-1)}(Z_2). \nonumber\eea
Solving (\ref{Box-rec-2}) we get the following final result for the recursion and reduction relation.  To use this formula, it is of course necessary to understand the cut box integrals as being defined in terms of the underlying arguments $K,P_1,P_2$, so that it is possible to find the necessary $Z_1,Z_2$ for reduction relations.
\bea & & {\rm Box}^{(n)}(A,B,C,D) =  F_{4\to 4}^{(n)}(A,B){\rm
Box}^{(0)}(A,B,C,D)+ \W F_{4\to
3}^{(n)}(A,B,C,D;Z_1){\rm Tri}^{(0)}(Z_1)\nonumber\\
& & ~~~~+ \W F_{4\to 3}^{(n)}(A,B,C,D;Z_2) {\rm Tri}^{(0)}(Z_2)+ \W
F_{4\to 2}^{(n)}(A,B,C,D;Z_i) {\rm Bub}^{(0)},~~~\label{I-4n-rec-2}
\eea
where these form factors are given by
\bea F_{4\to 4}^{(n)}(A,B) & = &
{\Gamma(1/2-\eps)\Gamma(n-\eps) \over
\Gamma(-\eps)\Gamma(n+1/2-\eps)} \left({B \over A}\right)^n,
~~~\label{F-4n-4n}
\\
\W F_{4\to 3}^{(n)}(A,B,C,D;Z_i) & = &-{\Gamma(n-\eps) \over
\Gamma(n+1/2-\eps)} {C_{Z_i}\over A ~Z_i} \sum_{k=1}^n
{\Gamma(k-1/2-\epsilon) \over \Gamma(k-1-\eps)}\left({B \over
A}\right)^{n-k} F^{(k-1)}_{3\to 3}(Z_i), ~~~\label{F-4n-3n}
\\
\W F_{4\to 2}^{(n)}(A,B,C,D;Z_i)  & = &-
 {\Gamma(n-\eps) \over \Gamma(n+1/2-\eps)}
{1\over A} \nonumber \\
& & \hspace{-1.2cm} \times
 \sum_{k=1}^n  {\Gamma(k-1/2-\epsilon) \over \Gamma(k-1-\eps)}\left({B \over A}\right)^{n-k}
\left( {C_{Z_1}\over Z_1} \W F^{(k-1)}_{3\to 2}(Z_1) + {C_{Z_2}\over
Z_2} \W F^{(k-1)}_{3\to 2}(Z_2) \right). ~~~\label{F-4n-2n} \eea

Again (\ref{I-4n-rec-2}) is not the final formula we are after. To
get the proper physical result, we need to replace the kinematic
factor ${1/ 2K^2}$. The result is
\bea & & \int_0^1 du~ u^{-1-\eps} u^n \left[{1\over 2K^2 \sqrt{ B -
A u}} \ln \left( {D - C u+ \sqrt{1-u}\sqrt{ B - A u}\over D - C u-
\sqrt{1-u}\sqrt{ B - A u}}\right)\right]  =  F_{4\to 4}^{(n)}(A,B)
C[I_4(K;P_1, P_2)] \nonumber \\ & & ~~~~~~~~~
+  \sum_{i=1}^2 F_{4\to
3}^{(n)}(A,B,C,D;Z_i) C[I_{3}(K_1^{(i)},K_3^{(i)})]+ F_{4\to
2}^{(n)}(A,B,C,D;Z_i) C[I_2(K)], ~~~\label{I-4n-rec-3} \eea
where for the triangles, $K_1^{(i)}$ and $K_3^{(i)}$ are given by Table
(\ref{Box-cut-table}), and the form factors are
\bea F_{4\to 3}^{(n)}(A,B,C,D;Z_i)& = & -{\sqrt{\Delta_{3}^{(i)}}\over
  2K^2} ~ \W
F_{4\to 3}^{(n)}(A,B,C,D;Z_i), \nonumber \\ F_{4\to
2}^{(n)}(A,B,C,D;Z_i)& = &  {1\over
2K^2} ~ \W F_{4\to 2}^{(n)}(A,B,C,D;Z_i).
\eea
%

\subsection{Cut pentagons}

The double cut will be
\bean {\delta(\W\ell^2-\mu^2) \delta( (\W\ell-K)^2-\mu^2) \over
((\W\ell-P_1)^2-\mu^2) ((\W\ell-P_2)^2-\mu^2)((\W\ell-P_3)^2-\mu^2)}.
\eean
Using $\W \ell=\ell+z K$ and doing the $t$-integration we reach
\bea C[I_5(K;P_1,P_2,P_3)]=\int_0^1 du~ u^{-1-\eps} \int
\vev{\ell~d\ell}[\ell~d\ell] {(1-2z) \over (K^2)^2}
{\gb{\ell|K|\ell} \over \gb{\ell|Q_1 |\ell} \gb{\ell|Q_2
|\ell}\gb{\ell|Q_3 |\ell}},
\eea
with
\bea Q_i =-(1-2z) P_i +{ P_i^2-2z P_i \cdot K \over K^2} K.
\eea
Now we do the splitting and get
\bean C[I_5(K;P_1,P_2,P_3)]& = &\int_0^1 du~ u^{-1-\eps} \int
\vev{\ell~d\ell}[\ell~d\ell] {(1-2z) \over (K^2)^2}( I_1 + I_2 + I_3) \\
I_1 & = & {\vev{\ell|K Q_1|\ell}^2 \over \vev{\ell|Q_2 Q_1|\ell}
\vev{\ell|Q_3 Q_1|\ell}} {1\over \gb{\ell|K|\ell} \gb{\ell|Q_1
|\ell}} \\ I_2 & = & -{\vev{\ell|K Q_2|\ell}^2 \over \vev{\ell|Q_2
Q_1|\ell} \vev{\ell|Q_3 Q_2|\ell}} {1\over \gb{\ell|K|\ell}
\gb{\ell|Q_2 |\ell}} \\
I_3 & = & {\vev{\ell|K Q_3|\ell}^2 \over \vev{\ell|Q_3 Q_2|\ell}
\vev{\ell|Q_3 Q_1|\ell}} {1\over \gb{\ell|K|\ell} \gb{\ell|Q_3
|\ell}}.
 \eean

 First we use
Feynman parametrization and then write  the integrand as a total
derivative \cite{Cachazo:2004by}. Next we do the Feynman parameter integration and finally
we read out the pole contribution.\footnote{In general we need to be
careful about  changing the order of taking residues and
doing the Feynman parameter integration. One can check that in the case
here, it is legitimate to do the Feynman parameter integral first.}

The general integration has been done in Appendix C (equation
(\ref{rep-rul-1})).  The result can be summarized as the following
replacement:
\bean {1\over \gb{\ell|K|\ell} \gb{\ell|Q |\ell}} \to -{1\over
\vev{\ell|Q K|\ell}}\ln \left( -{x_+\gb{\ell|K|\ell}\over
\gb{\ell|Q|\ell}}\right),
\eean
where $x_+$ is one solution to the equation $(Q+xK)^2=0$.
First, notice that after summing up residues of
all poles, the term with $\ln(-x_+)$ will not contribute, because
the sum of all residues of a holomorphic function is zero.
After dropping it we have
\bean C[I_5(K;P_1,P_2,P_3)]& = &\int_0^1 du~ u^{-1-\eps}
 {(1-2z) \over (K^2)^2} \left. ( I_1 + I_2 + I_3)\right|_{residue}, \\
I_1 & = & {\vev{\ell|K Q_1|\ell} \over \vev{\ell|Q_2 Q_1|\ell}
\vev{\ell|Q_3 Q_1|\ell}} \ln \left( {\gb{\ell|K|\ell}\over
\gb{\ell|Q_1|\ell}}\right), \\ I_2 & = & -{\vev{\ell|K Q_2|\ell} \over
\vev{\ell|Q_2 Q_1|\ell} \vev{\ell|Q_3 Q_2|\ell}} \ln \left(
{\gb{\ell|K|\ell}\over \gb{\ell|Q_2|\ell}}\right), \\
I_3 & = & {\vev{\ell|K Q_3|\ell} \over \vev{\ell|Q_3 Q_2|\ell}
\vev{\ell|Q_3 Q_1|\ell}} \ln \left( {\gb{\ell|K|\ell}\over
\gb{\ell|Q_3|\ell}}\right), \\\eean
where $|_{residue}$ means that we take residues of the two poles in $\vev{\ell|Q_i
Q_j|\ell}$.  This type of pole is discussed in detail in Appendix C.
Here, we apply the trick given by (\ref{K1K3-x}).
More concretely,  we have
\bea \vev{\ell|Q_i Q_j|\ell}=
{\vev{\ell~P_{ij,+}}\vev{\ell~P_{ij,-}}[P_{ij,+}~P_{ij,-}] \over
x_{ij,+}-x_{ij,-}},~~~~~x_{ij,\pm} ={- Q_i \cdot Q_j\pm \sqrt{
(Q_i\cdot Q_j)^2 -Q_i^2 Q_j^2} \over  Q_j^2}.~~~\label{xij-QQ}\eea
Thus after taking  the residue,  we will have the following replacement
\bea {1\over \vev{\ell|Q_i Q_j|\ell}}\to {(x_{ij,+}-x_{ij,-})\over
[P_{ij,+}~P_{ij,-}]\vev{P_{ij,+}~P_{ij,-}}} =-{ 1\over
2\sqrt{(Q_i\cdot Q_j)^2 -Q_i^2 Q_j^2}}.\eea

Now, applying it to the poles from $\vev{\ell|Q_2 Q_1|\ell}$, we need to add
the two contributions from $I_1$ and $I_2$. Defining
\bean F_{12}(\ell)={\vev{\ell|K Q_1|\ell} \over  \vev{\ell|Q_3
Q_1|\ell}},~~~~~ \W F_{12}(\ell)=-{\vev{\ell|K Q_2|\ell} \over
 \vev{\ell|Q_3 Q_2|\ell}},~~~~~\eean
it is easy to check that $F_{12} (P_{12}) + \W F_{12}(P_{12})=0$.
This identity is important for the cancellation of unphysical
singularities.

The sum of the $I_1$ and $I_2$ contributions
for the poles from $\vev{\ell|Q_2 Q_1|\ell}$ is given by
\bean & & -{ 1\over 2\sqrt{(Q_2\cdot Q_1)^2 -Q_2^2 Q_1^2}}
\left\{ {F_{12}(P_{12,+})+ F_{12}(P_{12,-})\over
2}\ln\left({\gb{P_{12,+}|K |P_{12,+}}\over
\gb{P_{12,+}|Q_1|P_{12,+}}} {\gb{P_{12,-}|Q_1|P_{12,-}}\over
\gb{P_{12,-}|K |P_{12,-}}}\right) \right.\\
& &  + {F_{12}(P_{12,+})- F_{12}(P_{12,-})\over
2}\ln\left({\gb{P_{12,+}|K |P_{12,+}}\over
\gb{P_{12,+}|Q_1|P_{12,+}}} {\gb{P_{12,-}|K |P_{12,-}}\over
\gb{P_{12,-}|Q_1|P_{12,-}}}\right)  \\ & & + {\W
F_{12}(P_{12,+})+ \W F_{12}(P_{12,-})\over 2}\ln\left({\gb{P_{12,+}|K
|P_{12,+}}\over \gb{P_{12,+}|Q_2|P_{12,+}}} { \gb{P_{12,-}|Q_2|P_{12,-}} \over \gb{P_{12,-}|K
|P_{12,-}}
}\right) \\
& & \left. + {\W F_{12}(P_{12,+})- \W F_{12}(P_{12,-})\over
2}\ln\left({\gb{P_{12,+}|K |P_{12,+}}\over
\gb{P_{12,+}|Q_2|P_{12,+}}} {\gb{P_{12,-}|K |P_{12,-}}\over
\gb{P_{12,-}|Q_2|P_{12,-}}}\right) \right\}.
\eean
Using the relation $F_{12} (P_{12}) + \W F_{12}(P_{12})=0$ we can
combine the first term and third term as well as second term and
fourth term. Next we use (\ref{Multiply}) and (\ref{Division}) to
get the final result
\bean & & -{ 1\over 2\sqrt{(Q_2\cdot Q_1)^2 -Q_2^2 Q_1^2}}
\left({F_{12}(P_{12,+})+ F_{12}(P_{12,-})\over 2}\ln{  Q_2 \cdot
Q_1- \sqrt{  (Q_2\cdot Q_1)^2 -Q_2^2 Q_1^2}\over  Q_2 \cdot
Q_1+ \sqrt{  (Q_2\cdot Q_1)^2 -Q_2^2 Q_1^2}} \right.\\
& & \left. + {F_{12}(P_{12,+})- F_{12}(P_{12,-})\over 2}\ln\left(
{Q_2^2\over Q_1^2} \right)\right),
\eean
where the first term has the right physical singularity while the second
term does not. However, we can rewrite the second term as
\bean -{ 1\over 2\sqrt{(Q_2\cdot Q_1)^2 -Q_2^2 Q_1^2}}\left(
{F_{12}(P_{12,+})- F_{12}(P_{12,-})\over 2}\ln (Q_1^2)^{-1}+{\W
F_{12}(P_{12,+})- \W F_{12}(P_{12,-})\over 2}\ln
(Q_2^2)^{-1}\right).
\eean
Now we recognize that it is the residue of the following
expression:
\bean  \left.\left({\vev{\ell|K Q_1|\ell} \over \vev{\ell|Q_2 Q_1|\ell}
\vev{\ell|Q_3 Q_1|\ell}} \ln (Q_1^2)^{-1} - {\vev{\ell|K Q_2|\ell}
\over \vev{\ell|Q_2 Q_1|\ell} \vev{\ell|Q_3 Q_2|\ell}} \ln
(Q_2^2)^{-1}\right)\right|_{residues~from~\vev{\ell|Q_2 Q_1|\ell}}.
\eean
Because it is a holomorphic function, when we sum up all residues we
get zero. This illustrates how the unphysical
singularities cancel out in the final result, so that we are left
with only the physical cut structure.

By similar manipulations we identify  the pole contribution of $\vev{\ell|Q_3 Q_1|\ell}$
as
\bean & & -{ 1\over 2\sqrt{(Q_3\cdot Q_1)^2 -Q_3^2 Q_1^2}}
{F_{13}(P_{13,+})+ F_{13}(P_{13,-})\over 2}\ln{  Q_3 \cdot Q_1-
\sqrt{  (Q_3\cdot Q_1)^2 -Q_3^2 Q_1^2}\over  Q_3 \cdot Q_1+
\sqrt{  (Q_3\cdot Q_1)^2 -Q_3^2 Q_1^2}}, \eean
and the contribution of $\vev{\ell|Q_3 Q_2|\ell}$ as
\bean & & -{ 1\over 2\sqrt{(Q_3\cdot Q_2)^2 -Q_3^2 Q_2^2}}
{F_{32}(P_{32,+})+ F_{32}(P_{32,-})\over 2}\ln{  Q_3 \cdot Q_2-
\sqrt{  (Q_3\cdot Q_2)^2 -Q_3^2 Q_2^2}\over  Q_3 \cdot Q_2+
\sqrt{  (Q_3\cdot Q_2)^2 -Q_3^2 Q_2^2}}, \eean
 where
\bean F_{13}(\ell)={\vev{\ell|K Q_1|\ell} \over  \vev{\ell|Q_2
Q_1|\ell}},~~~~~  F_{32}(\ell)=-{\vev{\ell|K Q_2|\ell} \over
 \vev{\ell| Q_2 Q_1|\ell}}.\eean

Collecting everything together we have
\bean & &  C[I_5(K;P_1,P_2,P_3)] = -\int_0^1 du~ u^{-1-\eps}
 {(1-2z) \over (K^2)^2} \\ & &
 \left({ 1\over 2\sqrt{(Q_3\cdot Q_2)^2 -Q_3^2 Q_2^2}}
{F_{32}(P_{32,+})+ F_{32}(P_{32,-})\over 2}\ln{  Q_3 \cdot Q_2-
\sqrt{  (Q_3\cdot Q_2)^2 -Q_3^2 Q_2^2}\over  Q_3 \cdot Q_2+
\sqrt{  (Q_3\cdot Q_2)^2 -Q_3^2 Q_2^2}}\right. \\ & & + { 1\over
2\sqrt{(Q_3\cdot Q_1)^2 -Q_3^2 Q_1^2}} {F_{13}(P_{13,+})+
F_{13}(P_{13,-})\over 2}\ln{  Q_3 \cdot Q_1- \sqrt{ (Q_3\cdot
Q_1)^2 -Q_3^2 Q_1^2}\over  Q_3 \cdot Q_1+ \sqrt{  (Q_3\cdot
Q_1)^2 -Q_3^2 Q_1^2}}  \\ & & \left. + { 1\over 2\sqrt{(Q_2\cdot
Q_1)^2 -Q_2^2 Q_1^2}} {F_{12}(P_{12,+})+ F_{12}(P_{12,-})\over
2}\ln{  Q_2 \cdot Q_1- \sqrt{  (Q_2\cdot Q_1)^2 -Q_2^2
Q_1^2}\over  Q_2 \cdot Q_1+ \sqrt{  (Q_2\cdot Q_1)^2 -Q_2^2
Q_1^2}} \right).
\eean
To write the result in a good form, refer to the spinor algebra in
Appendix C. There, in equation (\ref{Func-S}), we define the following function:
\bea
& & S[Q_i, Q_j, Q_k,K]  =  {T_1\over T_2},~~~\label{Func-S-sec}
\eea
with
\bea
T_1 = -8 \det \left( \begin{array}{lcr} K \cdot Q_k & Q_i \cdot K & Q_j \cdot K\\
Q_i \cdot Q_k & Q_i^2 & Q_i \cdot Q_j \\ Q_j \cdot Q_k & Q_i \cdot Q_j &
Q_j^2
\end{array} \right); ~~~~
T_2 = -4 \det \left( \begin{array}{lcr} Q_k^2 & Q_i \cdot Q_k & Q_j \cdot Q_k\\
Q_i \cdot Q_k & Q_i^2 & Q_i \cdot Q_j \\ Q_j \cdot Q_k & Q_i \cdot Q_j &
Q_j^2
\end{array} \right).~~~~\label{T1-T2-sec}
\eea
Our result can then be written as
\bea C[I_5(K;P_1,P_2,P_3)] &=& -\int_0^1 du~ u^{-1-\eps}
 {\sqrt{1-u} \over (K^2)^2}~~~\label{Pentagon-gen} \\ &  &
\left({ S[Q_3, Q_2, Q_1,K]\over
4\sqrt{(Q_3\cdot Q_2)^2 -Q_3^2 Q_2^2}} \ln{  Q_3 \cdot Q_2 -
\sqrt{  (Q_3\cdot Q_2)^2 -Q_3^2 Q_2^2}\over  Q_3 \cdot Q_2 +
\sqrt{  (Q_3\cdot Q_2)^2 -Q_3^2 Q_2^2}}\right. \\ & & + { S[Q_3,
Q_1, Q_2,K]\over 4\sqrt{(Q_3\cdot Q_1)^2 -Q_3^2 Q_1^2}} \ln{
Q_3 \cdot Q_1 - \sqrt{  (Q_3\cdot Q_1)^2 -Q_3^2 Q_1^2}\over  Q_3
\cdot Q_1 + \sqrt{  (Q_3\cdot Q_1)^2 -Q_3^2 Q_1^2}}  \\ & &
\left.+ { S[Q_2, Q_1, Q_3,K]\over 4\sqrt{(Q_2\cdot Q_1)^2
-Q_2^2 Q_1^2}} \ln{  Q_2 \cdot Q_1 - \sqrt{  (Q_2\cdot Q_1)^2
-Q_2^2 Q_1^2}\over  Q_2 \cdot Q_1 + \sqrt{ (Q_2\cdot Q_1)^2
-Q_2^2 Q_1^2}}\right).
\nonumber \eea
It is important to notice that $C[I_5]$ is given by three different
box cuts $C[I_4]$ multiplying corresponding factors ${S[\bullet]\over
2K^2}$. Thus the factor $S[\bullet]$, especially its denominator
$T_2$, which is the same for all three $S[\bullet]$,  can be considered
as another signature of a pentagon.

The reduction of a pentagon is very easy. Suppose that  we find the integral
\bean
\int_0^1
du~ u^{-1-\eps} u^n \sum_{i=1}^3 {S[i]\over 2K^2} C[I_4^{(i)}].
\eean
Then what we need to do is expand ${u^n\over T_2}$ (where $T_2$ is
the denominator of $S[\bullet]$) in the form $f[u]+ {A\over T_2}$, where
$f[u]$ is a polynomial in $u$ and $A$ is constant in $u$.
That is, we write
\bean   \int_0^1 du~ u^{-1-\eps} u^n \sum_{i=1}^3 {S[i]\over 2K^2}
C[I_4^{(i)}] &=& \int_0^1 du~ u^{-1-\eps} u^n \sum_{i=1}^3 {T_1[i]\over
2K^2 T_2} C[I_4^{(i)}] \\ & = & \int_0^1 du~ u^{-1-\eps}
\sum_{i=1}^3 {f(u) T_1[i]\over 2K^2}C[I_4^{(i)}]+ A\sum_{i=1}^3
{S[i]\over 2K^2} C[I_4^{(i)}].
\eean
Then $A$ is the coefficient of the true pentagon, and ${f(u) T_1[i] /
2K^2}$ are the reductive coefficients of the corresponding boxes.

\section{Examples}

Here we present a few basic examples to illustrate the main points of
our technique.  We begin with the case of five gluons of positive
helicity.  Later we list the $u$-integrals for four-gluon amplitudes,
just to show the structures that arise.  These amplitudes were first
computed to all orders in $\eps$ in \cite{Bern:1995db}.

\subsection{Five gluons of positive helicity}

In this section, we demonstrate the cut integration for the all-plus helicity
configuration of the five-gluon amplitude.  All cuts are of course trivially
related by permutation symmetry.  Here we work with the cut $C_{12}$.  We
show the calculation in some detail to illustrate our method.

The cut momentum is $K_{12}$, and we begin with the integrand
\bean I & = & 2 A_L((-\ell_1), 1,2,(-\ell_2)) A_R
(\ell_2,3,4,5,\ell_1) \\ & = & 2 {\mu^2 [1~2]\over
\vev{1~2}((-\ell_1+k_1)^2-\mu^2)} { \mu^2 [5|K_{345} \ell_2|3]\over
\vev{3~4}\vev{4~5} ((\ell_2+k_3)^2-\mu^2)
((\ell_1+k_5)^2-\mu^2)}.
\eean
The letter $I$ will actually represent the full integral; we neglect to write
the integral signs and measures while we follow the steps.
Notice that with our choice of direction of the propagator momentum, we have
$\ell_2=K_{12}-\ell_1$,
in keeping with the convention in (\ref{gen-var}).
After performing the $t$-integration and substituting
$\mu^2=z(1-z)s_{12}$, we have
\bea I & = & -{2 z^2(1-z)^2 (1-2z) [1~2]^2\over
\vev{3~4}\vev{4~5}} (I_1 + I_2 ), \\
I_1 & = & { z[5~3] \gb{\ell|K_{12}|\ell} \over
\gb{\ell|Q_1|\ell}\gb{\ell|Q_2|\ell} \gb{\ell|Q_3|\ell}},
~~~\label{5posi-I1-1}\\
I_2 & = & -{(1-2z) \gb{\ell|K_{12}|3}[\ell~5]\over
\gb{\ell|Q_1|\ell}\gb{\ell|Q_2|\ell}
\gb{\ell|Q_3|\ell}},~~~\label{5posi-I1-2} \eea
with
\bea Q_1 & = & (1-2z) k_1+ z K_{12}, \\ Q_2 &=& (1-2z) k_5+{ 2z
(K_{12}\cdot k_5) \over K_{12}^2} K_{12}, \\
Q_3 & = & -(1-2z) k_3+{ (1-z) 2 k_3\cdot
K_{12}\over K_{12}^2 }K_{12}. \eea
After
splitting the denominator factors with partial fractions, we can write the integral as a sum of three terms, related to one another by permuting $Q_1,Q_2,Q_3$.
\bean I & = & P_1+ P_2+ P_3, \\P_1 & = & -{ u^2\sqrt{1-u} [1~2]^2\over 8 \vev{3~4}\vev{4~5}} { \vev{\ell|K_{12}
Q_1|\ell} ( z[5~3] \vev{\ell|K_{12} Q_1|\ell}+(1-2z)
\gb{\ell|K_{12}|3}\gb{\ell|Q_1|5}) \over \vev{\ell|Q_2
Q_1|\ell}\vev{\ell|Q_3 Q_1|\ell}}{1\over
\gb{\ell|K_{12}|\ell}\gb{\ell|Q_1|\ell}},\\P_2 & = & { u^2\sqrt{1-u}
[1~2]^2\over 8 \vev{3~4}\vev{4~5}} {
\vev{\ell|K_{12} Q_2|\ell} ( z[5~3] \vev{\ell|K_{12}
Q_2|\ell}+(1-2z) \gb{\ell|K_{12}|3}\gb{\ell|Q_2|5}) \over
\vev{\ell|Q_2 Q_1|\ell}\vev{\ell|Q_3 Q_2|\ell}}{1\over
\gb{\ell|K_{12}|\ell}\gb{\ell|Q_2|\ell}},\\P_3 & = & -{ u^2\sqrt{1-u}
[1~2]^2\over 8 \vev{3~4}\vev{4~5}} {
\vev{\ell|K_{12} Q_3|\ell} ( z[5~3] \vev{\ell|K_{12}
Q_3|\ell}+(1-2z) \gb{\ell|K_{12}|3}\gb{\ell|Q_3|5}) \over
\vev{\ell|Q_3 Q_1|\ell}\vev{\ell|Q_3 Q_2|\ell}}{1\over
\gb{\ell|K_{12}|\ell}\gb{\ell|Q_3|\ell}}.\eean
%

\subsubsection{Spinor integration}

Let us start with $P_1$.  Upon writing it as a total derivative and
choosing the auxiliary spinor to be $\la_1$, we get
\bean P_1 & = & -\int_0^1 dx { u^2(1-u) [1~2]^2\over 8
\vev{3~4}\vev{4~5}} {  ( z[5~3] \vev{\ell|K_{12}
Q_1|\ell}+(1-2z) \gb{\ell|K_{12}|3}\gb{\ell|Q_1|5}) \over
\vev{\ell|Q_2 Q_1|\ell}\vev{\ell|Q_3 Q_1|\ell}}{
[1~\ell]\vev{1~\ell}\over \gb{\ell|R_1|\ell}(x z+1-x)},
\eean
with
\bean R_1= x Q_1+(1-x) K_{12}.
 \eean

There are four single poles from the factors $\vev{\ell|Q_2 Q_1|\ell}$
and $\vev{\ell|Q_3 Q_1|\ell}$. We can do the $x$-integration first.
Then we get
\bean P_1 & = &  -{ u^2(1-u) [1~2]^2\over 8
\vev{3~4}\vev{4~5}} {  ( z[5~3] \vev{\ell|K_{12}
Q_1|\ell}+(1-2z) \gb{\ell|K_{12}|3}\gb{\ell|Q_1|5}) \over
\vev{\ell|Q_2 Q_1|\ell}\vev{\ell|Q_3 Q_1|\ell}}{
[1~\ell]\vev{1~\ell}\over \sqrt{1-u} \gb{\ell|k_1|\ell}}\left( \ln
(z)-\ln {\gb{\ell|Q_1|\ell}\over \gb{\ell|K_{12}|\ell}}\right)\\& =
& { u^2\sqrt{1-u} [1~2] K_{12}^2\over 8 \vev{1~2}\vev{3~4}\vev{4~5}}
{ ( z[5~3] \vev{\ell|K_{12} Q_1|\ell}+(1-2z)
\gb{\ell|K_{12}|3}\gb{\ell|Q_1|5}) \over \vev{\ell|Q_2
Q_1|\ell}\vev{\ell|Q_3 Q_1|\ell}}\ln {z\gb{\ell|K_{12}|\ell}
\over \gb{\ell|Q_1|\ell}}.\\
\eean

The next thing is to read out the residues of these four single poles.
Notice that since we can write
\bean
\ln {z\gb{\ell|K_{12}|\ell} \over
\gb{\ell|Q_1|\ell}}=\ln z+\ln {\gb{\ell|K_{12}|\ell} \over
\gb{\ell|Q_1|\ell}},
\eean
 and since the sum of residues of a holomorphic
function is zero, we get
\bean P_1 & = & \left.\left({ u^2\sqrt{1-u} [1~2]^2\over 8
\vev{3~4}\vev{4~5}} { ( z[5~3] \vev{\ell|K_{12}
Q_1|\ell}+(1-2z) \gb{\ell|K_{12}|3}\gb{\ell|Q_1|5}) \over
\vev{\ell|Q_2 Q_1|\ell}\vev{\ell|Q_3 Q_1|\ell}}\ln
{\gb{\ell|K_{12}|\ell} \over
\gb{\ell|Q_1|\ell}}\right)\right|_{residues}.\eean
Similar calculations give
\bean P_2 & = & \left.\left(-{ u^2\sqrt{1-u} [1~2]^2\over 8
\vev{3~4}\vev{4~5}} {  ( z[5~3] \vev{\ell|K_{12}
Q_2|\ell}+(1-2z) \gb{\ell|K_{12}|3}\gb{\ell|Q_2|5}) \over
\vev{\ell|Q_2 Q_1|\ell}\vev{\ell|Q_3 Q_2|\ell}} \ln  {
\gb{\ell|K_{12}|\ell}\over
\gb{\ell|Q_2|\ell}}\right)\right|_{residues}\eean
and
\bean P_3 & = & \left.\left({ u^2\sqrt{1-u} [1~2]^2\over 8
\vev{3~4}\vev{4~5}} {  ( z[5~3] \vev{\ell|K_{12}
Q_3|\ell}+(1-2z) \gb{\ell|K_{12}|3}\gb{\ell|Q_3|5}) \over
\vev{\ell|Q_3 Q_1|\ell}\vev{\ell|Q_3 Q_2|\ell}}\ln  {
\gb{\ell|K_{12}|\ell}\over\gb{\ell|Q_3|\ell}} \right)\right|_{residues}.
\eean
%

\subsubsection{Taking the residues}

Now we need to take the residues of $P_i$. Again we start with $P_1$.
The pole contribution of $\vev{\ell|Q_2 Q_1|\ell}$ can  be
written as
\bean {\vev{1~2}^2\over \vev{2|Q_2 Q_1 |2}} {1\over
\vev{\eta_1~\eta_2}}\left( F_1(\eta_1) \ln
{\gb{\eta_1|K_{12}|\eta_1}\over \gb{\eta_1|Q_1|\eta_1}}-F_1(\eta_2)
\ln {\gb{\eta_2|K_{12}|\eta_2}\over
\gb{\eta_2|Q_1|\eta_2}}\right),
\eean
where $\eta_1,\eta_2$ are the two solutions of $\vev{\ell|Q_2 Q_1|\ell}=0$
(see Appendix C),  and $F_1(\eta)$ is defined as
\bea F_1(\eta)={ ( z[5~3] \vev{\eta|K_{12} Q_1|\eta}+(1-2z)
\gb{\eta|K_{12}|3}\gb{\eta|Q_1|5}) \over \vev{\eta|Q_3 Q_1|\eta}}.
\eea
Decompose $F(\eta)$ into two pieces that are respectively symmetric and antisymmetric under the exchange $\eta_1\leftrightarrow
\eta_2$:
\bea F_1(\eta_1)=F_1^S+F_1^A,~~~~F_1(\eta_2)=F_1^S-F_1^A,\eea
and the pole contribution can be written as
\bean {\vev{1~2}^2\over \vev{2|Q_2 Q_1 |2}} {1\over
\vev{\eta_1~\eta_2}}\left( F_1^S\ln
{\gb{\eta_1|K_{12}|\eta_1}\over
\gb{\eta_1|Q_1|\eta_1}}
{\gb{\eta_2|Q_1|\eta_2}\over \gb{\eta_2|K_{12}|\eta_2}}
+ F_1^A \ln {{\gb{\eta_1|K_{12}|\eta_1}\over
\gb{\eta_1|Q_1|\eta_1}} {\gb{\eta_2|K_{12}|\eta_2}\over
\gb{\eta_2|Q_1|\eta_2}}} \right).
\eean
We substitute the solutions $\eta_1,\eta_2$ and define
\bea
\W A \equiv
{s_{25} \over s_{15}}.
\eea
Then we find
\bean & & {\vev{1~2}^2\over \vev{2|Q_2 Q_1 |2}} {1\over
\vev{\eta_1~\eta_2}}\left( F_1^S\ln {\sqrt{1+\W A u}-\sqrt{1-u}\over
\sqrt{1+\W A u}+\sqrt{1-u}}+F_1^A \ln { 4(1+\W A)\over u(1+\W A u)}
\right)
\\ &= & {\vev{1~2}^2\over \vev{2|Q_2 Q_1 |2}} {1\over
\vev{\eta_1~\eta_2}}\left( {F_1(\eta_1)+F_1(\eta_2)\over 2}\ln
{\sqrt{1+\W A u}-\sqrt{1-u}\over \sqrt{1+\W A
u}+\sqrt{1-u}}+{F_1(\eta_1)-F_1(\eta_2)\over 2} \ln { 4(1+\W A)\over
u(1+\W A u)} \right).
\eean
 Notice that the second term
 can be interpreted as the pole
contribution of $\vev{\ell|Q_2 Q_1|\ell}$. This observation will be
useful to prove that the sum of all pole contributions is zero.

Now we consider the contribution of $\vev{\ell|Q_3 Q_1|\ell}$. Using
$\eta_3, \eta_4$ as the solutions with
\bea
\W B \equiv \left( {s_{13} \over
s_{23}}\right)^2,
\eea
 and the definition
\bea F_2(\eta) \equiv {  z[5~3] \vev{\eta|K_{12} Q_1|\eta}+(1-2z)
\gb{\eta|K_{12}|3}\gb{\eta|Q_1|5} \over \vev{\eta|Q_2 Q_1|\eta}},
\eea
we get
\bean {\vev{1~2}^2\over \vev{2|Q_3 Q_1 |2}} {1\over
\vev{\eta_3~\eta_4}}\left( {F_2(\eta_3)+F_2(\eta_4)\over 2}\ln
{\sqrt{1+\W B  u}-\sqrt{1-u}\over \sqrt{1+\W B
u}+\sqrt{1-u}}+{F_2(\eta_3)-F_2(\eta_4)\over 2} \ln { 4(1+\W B
)\over u(1+\W B  u)} \right).
\eean
Putting everything together, we have
\bean P_1 & = & { u^2\sqrt{1-u} [1~2]^2\over 8
\vev{3~4}\vev{4~5}} \\ & & \left({\vev{1~2}^2\over
\vev{2|Q_2 Q_1 |2}} {1\over \vev{\eta_1~\eta_2}}\left(
{F_1(\eta_1)+F_1(\eta_2)\over 2}\ln {\sqrt{1+\W A u}-\sqrt{1-u}\over
\sqrt{1+\W A u}+\sqrt{1-u}}+{F_1(\eta_1)-F_1(\eta_2)\over 2} \ln {
4(1+\W A)\over u(1+\W A u)} \right)\right. \\ & & \left.
+{\vev{1~2}^2\over \vev{2|Q_3 Q_1 |2}} {1\over
\vev{\eta_3~\eta_4}}\left( {F_2(\eta_3)+F_2(\eta_4)\over 2}\ln
{\sqrt{1+\W B  u}-\sqrt{1-u}\over \sqrt{1+\W B
u}+\sqrt{1-u}}+{F_2(\eta_3)-F_2(\eta_4)\over 2} \ln { 4(1+\W B
)\over u(1+\W B  u)} \right)\right).
\eean
Let us examine the various terms in this expression.
Some of the singularities are spurious.
 First, the terms with  $\ln
{4\over u}$ can be written as
\bean { u^2\sqrt{1-u} [1~2]^2\over 8
\vev{3~4}\vev{4~5}}\ln {4\over u}\left.\left({  z[5~3]
\vev{\ell|K_{12} Q_1|\ell}+(1-2z) \gb{\ell|K_{12}|3}\gb{\ell|Q_1|5}
\over \vev{\ell|Q_2 Q_1|\ell}\vev{\ell|Q_3
Q_1|\ell}}\right)\right|_{residues}.
\eean
Since the function is holomorphic, the
sum of the residues of all poles will be zero, so these terms can be
discarded collectively.

Then we are left with 
\bea P_1 & = & { u^2\sqrt{1-u} [1~2]^2\over 8
\vev{3~4}\vev{4~5}} \left({\vev{1~2}^2\over \vev{2|Q_2 Q_1
|2}} {1\over \vev{\eta_1~\eta_2}} {F_1(\eta_1)+F_1(\eta_2)\over
2}\ln {\sqrt{1+\W A u}-\sqrt{1-u}\over \sqrt{1+\W A
u}+\sqrt{1-u}}\right.\nonumber \\ & & +{\vev{1~2}^2\over \vev{2|Q_3
Q_1 |2}} {1\over \vev{\eta_3~\eta_4}} {F_2(\eta_3)+F_2(\eta_4)\over
2}\ln {\sqrt{1+\W B  u}-\sqrt{1-u}\over \sqrt{1+\W B
u}+\sqrt{1-u}}\nonumber \\ & & \left. + {1\over 2}\left.\left({  z[5~3]
\vev{\ell|K_{12} Q_1|\ell}+(1-2z) \gb{\ell|K_{12}|3}\gb{\ell|Q_1|5}
\over \vev{\ell|Q_3 Q_1|\ell}\vev{\ell|Q_2
Q_1|\ell}}\right)\right|_{\vev{\ell|Q_2 Q_1|\ell}=0}\ln \left({ 1+\W
A\over 1+\W A u}\right)\right. \nonumber \\  & & \left. + {1\over
2}\left.\left({  z[5~3] \vev{\ell|K_{12} Q_1|\ell}+(1-2z)
\gb{\ell|K_{12}|3}\gb{\ell|Q_1|5} \over \vev{\ell|Q_3
Q_1|\ell}\vev{\ell|Q_2 Q_1|\ell}}\right)\right|_{\vev{\ell|Q_3
Q_1|\ell}=0}\ln \left({ 1+\W B \over 1+\W B
u}\right)\right).~~~~\label{P1-pole}\eea
In this expression we have written the last two terms 
in a form where the poles need to be substituted.

Similarly we have
\bea P_2 & = & - { u^2\sqrt{1-u} [1~2]^2\over 8
\vev{3~4}\vev{4~5}}\left( {\vev{3~5}^2\over \vev{5|Q_3
Q_2|5}}{1\over \vev{\eta_5~\eta_6}} {F_4(\eta_5)+F_4(\eta_6)\over
2}\ln {\sqrt{1+ \W C u}+\sqrt{1-u}\over \sqrt{1+ \W C u
}-\sqrt{1-u}}\right. \nonumber \\ & & +{\vev{1~2}^2\over \vev{2|Q_2
Q_1|2}} {1\over \vev{\eta_1~\eta_2}} { F_3(\eta_1)+F_3(\eta_2)\over
2}\ln {\sqrt{1+ \W A u}+\sqrt{1-u}\over \sqrt{1+ \W A
u}-\sqrt{1-u}}\nonumber \\ & & + {1\over 2}\left.\left({   z[5~3]
\vev{\ell|K_{12} Q_2|\ell}+(1-2z) \gb{\ell|K_{12}|3}\gb{\ell|Q_2|5}
\over \vev{\ell|Q_2 Q_1|\ell}\vev{\ell|Q_3
Q_2|\ell}}\right)\right|_{\vev{\ell|Q_3 Q_2|\ell}=0}\ln \left({ 1+\W C
\over 1+ \W C  u}\right)\nonumber \\& & \left.+ {1\over 2}\left.\left({
z[5~3] \vev{\ell|K_{12} Q_2|\ell}+(1-2z)
\gb{\ell|K_{12}|3}\gb{\ell|Q_2|5} \over \vev{\ell|Q_2
Q_1|\ell}\vev{\ell|Q_3 Q_2|\ell}}\right)\right|_{\vev{\ell|Q_2
Q_1|\ell}=0}\ln \left({1+\W A\over 1+\W A u}\right)\right),~~~\label{P2-pole}
\eea
where
\bea
\W C  \equiv { s_{12} s_{35} \over s_{34} s_{45} }
\eea
 and
\bea F_3(\eta)={  z[5~3] \vev{\eta|K_{12} Q_2|\eta}+(1-2z)
\gb{\eta|K_{12}|3}\gb{\eta|Q_2|5} \over \vev{\eta|Q_3 Q_2|\eta}},
\eea
\bea F_4(\eta)={  z[5~3] \vev{\eta|K_{12} Q_2|\eta}+(1-2z)
\gb{\eta|K_{12}|3}\gb{\eta|Q_2|5} \over \vev{\eta|Q_2 Q_1|\eta}}.
\eea
Similarly again for $P_3$, we have
\bea P_3 & = & { u^2\sqrt{1-u} [1~2]^2\over 8
\vev{3~4}\vev{4~5}} \left( {\vev{3~5}^2\over \vev{5|Q_3
Q_2|5}}{1\over \vev{\eta_5~\eta_6}} {F_6(\eta_5)+F_6(\eta_6)\over 2}
\ln {\sqrt{1+ \W C  u}-\sqrt{1-u}\over \sqrt{1+ \W C
u}+\sqrt{1-u}}\right.\nonumber \\ & & +{\vev{1~2}^2\over
\vev{\ell|Q_3 Q_1|\ell}}{1\over \vev{\eta_3~\eta_4}}
{F_5(\eta_3)+F_5(\eta_4)\over 2} \ln {\sqrt{1+ \W B u
}+\sqrt{1-u}\over \sqrt{1+ \W B u }-\sqrt{1-u}}\nonumber  \\ & &
+{1\over 2}\left.\left({   z[5~3] \vev{\ell|K_{12} Q_3|\ell}+(1-2z)
\gb{\ell|K_{12}|3}\gb{\ell|Q_3|5} \over \vev{\ell|Q_3
Q_1|\ell}\vev{\ell|Q_3 Q_2|\ell}}\right)\right|_{\vev{\ell|Q_3
Q_1|\ell}=0}\ln \left({ 1+\W B \over 1+\W B  u}\right) \nonumber \\ & &
\left.+{1\over 2}\left.\left({   z[5~3] \vev{\ell|K_{12} Q_3|\ell}+(1-2z)
\gb{\ell|K_{12}|3}\gb{\ell|Q_3|5} \over \vev{\ell|Q_3
Q_1|\ell}\vev{\ell|Q_3 Q_2|\ell}}\right)\right|_{\vev{\ell|Q_3
Q_2|\ell}=0}\ln \left({ 1+\W C \over 1+\W C  u}\right)\right),~~\label{P3-pole}\eea
where
\bea F_5(\eta) \equiv {  z[5~3] \vev{\eta|K_{12} Q_3|\eta}+(1-2z)
\gb{\eta|K_{12}|3}\gb{\eta|Q_3|5} \over \vev{\eta|Q_3 Q_2|\eta}},
\eea
\bea F_6(\eta) \equiv {  z[5~3] \vev{\eta|K_{12} Q_3|\eta}+(1-2z)
\gb{\eta|K_{12}|3}\gb{\eta|Q_3|5} \over \vev{\eta|Q_3 Q_1|\eta}}.
\eea
%

\subsubsection{Summing up the result}

Now we sum up $P_1, P_2, P_3$. First we check that the
spurious singularities cancel out. For $\ln { 1+\W A\over 1+\W A u}$
we get
\bean & & \left({  z[5~3] \vev{\ell|K_{12} Q_1|\ell}+(1-2z)
\gb{\ell|K_{12}|3}\gb{\ell|Q_1|5} \over \vev{\ell|Q_3
Q_1|\ell}\vev{\ell|Q_2 Q_1|\ell}}\right)-\left({   z[5~3]
\vev{\ell|K_{12} Q_2|\ell}+(1-2z) \gb{\ell|K_{12}|3}\gb{\ell|Q_2|5}
\over \vev{\ell|Q_2 Q_1|\ell}\vev{\ell|Q_3 Q_2|\ell}}\right)\\
& = & { z[5~3]\vev{\ell|K_{12} Q_3|\ell}\vev{\ell|Q_1 Q_2|\ell}
+(1-2z) \gb{\ell|K_{12}|3} \gb{\ell|Q_3|5}\vev{\ell|Q_1
Q_2|\ell}\over \vev{\ell|Q_3 Q_1|\ell}\vev{\ell|Q_2
Q_1|\ell}\vev{\ell|Q_3 Q_2|\ell}},\eean
where we should calculate only the pole contribution from
$\vev{\ell|Q_2 Q_1|\ell}$. However, the factor $\vev{\ell|Q_1
Q_2|\ell}$ in the numerator shows us that the contribution is zero. Thus
the singularity in $\ln \left({ 1+\W A\over 1+\W A u}\right)$ disappears from the final
result.

For $\ln \left({ 1+\W B \over 1+\W B  u}\right)$, we have
\bean & & \left({  z[5~3] \vev{\ell|K_{12} Q_1|\ell}+(1-2z)
\gb{\ell|K_{12}|3}\gb{\ell|Q_1|5} \over \vev{\ell|Q_3
Q_1|\ell}\vev{\ell|Q_2 Q_1|\ell}}\right)+\left({   z[5~3]
\vev{\ell|K_{12} Q_3|\ell}+(1-2z) \gb{\ell|K_{12}|3}\gb{\ell|Q_3|5}
\over \vev{\ell|Q_3 Q_1|\ell}\vev{\ell|Q_3 Q_2|\ell}}\right)
\\ & = & {\vev{\ell|Q_3 Q_1|\ell}( z[5~3]\vev{\ell|K_{12} Q_2|\ell}
+(1-z)\gb{\ell|K_{12}|3}\gb{\ell|Q_2|5})\over \vev{\ell|Q_3
Q_1|\ell}\vev{\ell|Q_2 Q_1|\ell}\vev{\ell|Q_3 Q_2|\ell}}.
\eean
Again, the numerator factor $\vev{\ell|Q_3 Q_1|\ell}$ tells us the sum is zero.

For $\ln \left({ 1+\W C \over 1+\W C  u}\right)$, we have
\bean & &-\left({   z[5~3] \vev{\ell|K_{12} Q_2|\ell}+(1-2z)
\gb{\ell|K_{12}|3}\gb{\ell|Q_2|5} \over \vev{\ell|Q_2
Q_1|\ell}\vev{\ell|Q_3 Q_2|\ell}}\right) +\left({   z[5~3]
\vev{\ell|K_{12} Q_3|\ell}+(1-2z) \gb{\ell|K_{12}|3}\gb{\ell|Q_3|5}
\over \vev{\ell|Q_3 Q_1|\ell}\vev{\ell|Q_3 Q_2|\ell}}\right)
\\ & = & -{\vev{\ell|Q_3 Q_2|\ell}( z[5~3]\vev{\ell|K_{12} Q_1|\ell}
+(1-z)\gb{\ell|K_{12}|3}\gb{\ell|Q_1|5})\over \vev{\ell|Q_3
Q_1|\ell}\vev{\ell|Q_2 Q_1|\ell}\vev{\ell|Q_3 Q_2|\ell}}.
\eean
Again, the numerator factor $\vev{\ell|Q_3 Q_2|\ell}$ tells us the sum is zero.

Now we consider the remaining singularities. For the first factor we have
\bean I&= &  { u^2\sqrt{1-u} [1~2]^2\over 8
\vev{3~4}\vev{4~5}} {\vev{1~2}^2\over \vev{2|Q_2 Q_1 |2}}
{1\over \vev{\eta_1~\eta_2}}
{F_1(\eta_1)+F_1(\eta_2)+F_3(\eta_1)+F_3(\eta_2)\over 2}\ln
{\sqrt{1+\W A u}-\sqrt{1-u}\over \sqrt{1+\W A u}+\sqrt{1-u}}\\ & &
+{ u^2\sqrt{1-u} [1~2]^2\over 8
\vev{3~4}\vev{4~5}}{\vev{1~2}^2\over \vev{2|Q_3 Q_1 |2}}
{1\over \vev{\eta_3~\eta_4}}
{F_2(\eta_3)+F_2(\eta_4)-F_5(\eta_3)+F_5(\eta_4)\over 2}\ln
{\sqrt{1+\W B  u}-\sqrt{1-u}\over \sqrt{1+\W B  u}+\sqrt{1-u}} \\ &
& +{ u^2\sqrt{1-u} [1~2]^2\over 8
\vev{3~4}\vev{4~5}} {\vev{3~5}^2\over \vev{5|Q_3
Q_2|5}}{1\over \vev{\eta_5~\eta_6}}
 {F_6(\eta_5)+F_6(\eta_6)+F_4(\eta_5)+F_4(\eta_6)\over 2} \ln {\sqrt{1+
\W C  u}-\sqrt{1-u}\over \sqrt{1+ \W C  u}+\sqrt{1-u}}\\ & = & {
u^2\sqrt{1-u} [1~2]^2\over 8 \vev{3~4}\vev{4~5}}
{\vev{1~2}^2\over \vev{2|Q_2 Q_1 |2}} {1\over \vev{\eta_1~\eta_2}}
(F_1(\eta_1)+F_1(\eta_2))\ln {\sqrt{1+\W A u}-\sqrt{1-u}\over
\sqrt{1+\W A u}+\sqrt{1-u}}\\ & & +{ u^2\sqrt{1-u} [1~2]^2\over 8 \vev{3~4}\vev{4~5}}{\vev{1~2}^2\over
\vev{2|Q_3 Q_1 |2}} {1\over \vev{\eta_3~\eta_4}}
(F_2(\eta_3)+F_2(\eta_4))\ln {\sqrt{1+\W B  u}-\sqrt{1-u}\over
\sqrt{1+\W B  u}+\sqrt{1-u}} \\ & & +{ u^2\sqrt{1-u} [1~2]^2\over 8 \vev{3~4}\vev{4~5}} {\vev{3~5}^2\over
\vev{5|Q_3 Q_2|5}}{1\over \vev{\eta_5~\eta_6}}
 (F_4(\eta_5)+F_4(\eta_6)) \ln {\sqrt{1+
\W C  u}-\sqrt{1-u}\over \sqrt{1+ \W C  u}+\sqrt{1-u}}. \eean
It is easy to check that $F_1(\eta_{1,2})=F_3(\eta_{1,2})$ up to the term
$\vev{\ell|Q_2 Q_1|\ell}$ which is zero in our case. Similarly
$F_2(\eta_{3,4})=-F_5(\eta_{3,4})$ and
$F_4(\eta_{5,6})=F_6(\eta_{5,6})$. We need to carry out the
summation, especially to show that the factor $\sqrt{1-u}$ cancels out.

The summation can be carried out using the technique presented in
Appendix C, and we get
\bean I & = & {s_{12}^2\over 8
\vev{1~2}\vev{2~3}\vev{3~4}\vev{4~5}\vev{5~1}}
( u^2 T + u^3 \gb{2|k_3 k_1 k_5-k_5 k_1 k_3|2} U ) \\
T & \equiv &   {1\over \sqrt{1+ u \W A}}\ln {\sqrt{1+\W A
u}-\sqrt{1-u}\over \sqrt{1+\W A u}+\sqrt{1-u}}-{1\over \sqrt{1+ u \W
B }}\ln {\sqrt{1+\W B  u}-\sqrt{1-u}\over \sqrt{1+\W B
u}+\sqrt{1-u}}  \\ & & - {1\over \sqrt{1+ \W C  u}}\ln
{\sqrt{1+ \W C
u}-\sqrt{1-u}\over \sqrt{1+ \W C  u}+\sqrt{1-u}}\\
U & \equiv & {\gb{2|k_5 k_4 k_3+k_3 k_4 k_5 |2}\over 4 s_{51}
s_{23} s_{34} s_{45}- \gb{2|k_3 k_1 k_5-k_5 k_1 k_3|2}^2 u}
{1\over \sqrt{1+ u \W A}}\ln {\sqrt{1+\W A u}-\sqrt{1-u}\over
\sqrt{1+\W A u}+\sqrt{1-u}}
\\ & & -{\gb{3|k_4 k_5 k_1+ k_1 k_5 k_4|3}\over 4 s_{51} s_{23}
s_{34} s_{45}- \gb{2|k_3 k_1 k_5-k_5 k_1 k_3|2}^2 u}{1\over
\sqrt{1+ u \W B }}\ln {\sqrt{1+\W B  u}-\sqrt{1-u}\over \sqrt{1+\W B
u}+\sqrt{1-u}}
\\ & &  -{\gb{5|k_1 k_2
k_3+k_3 k_2 k_1|5}\over 4 s_{51} s_{23} s_{34} s_{45}-
\gb{2|k_3 k_1 k_5-k_5 k_1 k_3|2}^2 u}{1\over \sqrt{1+ \W C  u}}\ln
{\sqrt{1+ \W C  u}-\sqrt{1-u}\over \sqrt{1+ \W C
u}+\sqrt{1-u}}.
\eean

It is easy to see that $T$ is a pure box contribution and $U$ is the
exact expression for the pentagon. The coefficient $u^3$ in
front of $U$ is easy to deal with. Since
there is a common denominator factor in the three terms of $U$, we
write
\bea u^3\to \left( \left( u- {4 s_{51} s_{23} s_{34}
s_{45}\over \gb{2|k_3 k_1 k_5-k_5 k_1 k_3|2}^2}\right)+{4 s_{51}
s_{23} s_{34} s_{45}\over \gb{2|k_3 k_1 k_5-k_5 k_1
k_3|2}^2}\right)^3,
 \eea
and make the expansion. Some terms go to boxes, and the remainder is
the pure pentagon contribution.

To finish the program and  read out the exact coefficients, we need to 
identify the cut boxes exactly for this amplitude.  They are the
following.  (See also the subsection on one-mass boxes in Appendix B.)
\begin{itemize}

\item (1)  For box
$(12|3|4|5)$, we have   $K= K_{12}$, $P_1= -k_5$ and $P_2= -K_{45}$.
Thus $A=-{s_{45} s_{34} s_{35} \over 4 s_{12}}>0$, $D={
s_{34} s_{45} \over 2 s_{12}}>0$ and $B=D^2$.  Notice that here
$A,B,C,D$ are defined as in (\ref{Para-3}), and
the quantities $\W A, \W B, \W C$ we have defined in this section are just ${-A/ B}$ in
various cuts. Then we find
\bea C[I_{12|3|4|5}] & = & {2\over s_{34} s_{45}}\int_0^1 du
u^{-1-\eps} {1\over \sqrt{1+ \W C  u}}\ln {\sqrt{1+ \W C
u}-\sqrt{1-u}\over \sqrt{1+ \W C  u}+\sqrt{1-u}}. 
\eea 

\item (2) For $(1|2|3|45)$, we have $K=-K_{12}$, $P_1=k_3$ and $P_2=-k_2$, thus by
(\ref{Para-3}) we have $D=-{s_{23}\over 2}>0$, $A=-{ s_{23}
s_{13}\over 4}<0$ and $B=D^2$. Thus
\bea C[I_{1|2|3|45}] & = & {2\over s_{12} s_{23}}\int_0^1 du
u^{-1-\eps} {1\over \sqrt{1+ \W B u}}\ln {\sqrt{1+ \W B
u}-\sqrt{1-u}\over \sqrt{1+ \W B  u}+\sqrt{1-u}}.
\eea

\item (6) For $(1|2|34|5)$, we have $P_1=k_1$, $P_2=-k_5$ and
$K=K_{12}$, thus by (\ref{Para-3}) we have $D={s_{51}\over 2}>0$,
$A=-{s_{51} s_{52}\over 4}<0$ and $B=D^2$. Thus
\bea C[I_{1|2|34|5}] & = & -{2\over s_{12} s_{51}} \int_0^1 du
u^{-1-\eps} {1\over \sqrt{1+ A u}}\ln {\sqrt{1+ A u}-\sqrt{1-u}\over
\sqrt{1+ A u}+\sqrt{1-u}}.
\eea

\end{itemize}

Collecting all results, we find the following coefficients.
(These are the integrands for $\int_0^1 du u^{-1-\eps}$.)
\bean C_{pentagon} & = & -{s_{12}^3\gb{2|k_3 k_1 k_5-k_5 k_1
k_3|2}\over 32
\vev{1~2}\vev{2~3}\vev{3~4}\vev{4~5}\vev{5~1}}\left({4 s_{51}
s_{23} s_{34} s_{45}\over \gb{2|k_3 k_1 k_5-k_5 k_1
k_3|2}^2}\right)^3\\
C_{1|2|34|5} & = & -{ s_{12}^3 s_{51}\over 16
\vev{1~2}\vev{2~3}\vev{3~4}\vev{4~5}\vev{5~1}}\left( u^2- {\gb{2|k_5
k_4 k_3+k_3 k_4 k_5 |2}\over \gb{2|k_3 k_1 k_5-k_5 k_1
k_3|2}}\right. \\ & & \left. \left(\left( u- {4 s_{51} s_{23}
s_{34} s_{45}\over \gb{2|k_3 k_1 k_5-k_5 k_1
k_3|2}^2}\right)^2+3 \left( u- {4 s_{51} s_{23} s_{34}
s_{45}\over \gb{2|k_3 k_1 k_5-k_5 k_1 k_3|2}^2}\right){4 s_{51}
s_{23} s_{34} s_{45}\over \gb{2|k_3 k_1 k_5-k_5 k_1
k_3|2}^2}\right. \right.\\ & & \left. \left. +3 \left({4 s_{51}
s_{23} s_{34} s_{45}\over \gb{2|k_3 k_1 k_5-k_5 k_1
k_3|2}^2}\right)^2\right)\right)\\
C_{1|2|3|45} & = & -{s_{12}^3
s_{23}\over 16 \vev{1~2}\vev{2~3}\vev{3~4}\vev{4~5}\vev{5~1}}\left(
u^2- {\gb{3|k_4 k_5 k_1+ k_1 k_5 k_4|3}\over \gb{2|k_3 k_1 k_5-k_5
k_1 k_3|2}}\right. \\ & & \left. \left(\left( u- {4 s_{51}
s_{23} s_{34} s_{45}\over \gb{2|k_3 k_1 k_5-k_5 k_1
k_3|2}^2}\right)^2+3 \left( u- {4 s_{51} s_{23} s_{34}
s_{45}\over \gb{2|k_3 k_1 k_5-k_5 k_1 k_3|2}^2}\right){4 s_{51}
s_{23} s_{34} s_{45}\over \gb{2|k_3 k_1 k_5-k_5 k_1
k_3|2}^2}\right. \right.\\ & & \left. \left. +3 \left({4 s_{51}
s_{23} s_{34} s_{45}\over \gb{2|k_3 k_1 k_5-k_5 k_1
k_3|2}^2}\right)^2\right)\right)\\
C_{12|3|4|5} & = & -{s_{12}^2
s_{34} s_{45}\over 16
\vev{1~2}\vev{2~3}\vev{3~4}\vev{4~5}\vev{5~1}}\left( u^2- {\gb{5|k_1
k_2 k_3+k_3 k_2 k_1|5}\over \gb{2|k_3 k_1 k_5-k_5 k_1 k_3|2}}\right.
\\ & & \left. \left(\left( u- {4 s_{51} s_{23} s_{34}
s_{45}\over \gb{2|k_3 k_1 k_5-k_5 k_1 k_3|2}^2}\right)^2+3 \left(
u- {4 s_{51} s_{23} s_{34} s_{45}\over \gb{2|k_3 k_1 k_5-k_5
k_1 k_3|2}^2}\right){4 s_{51} s_{23} s_{34} s_{45}\over
\gb{2|k_3 k_1 k_5-k_5 k_1 k_3|2}^2}\right. \right.\\ & & \left.
\left. +3 \left({4 s_{51} s_{23} s_{34} s_{45}\over
\gb{2|k_3 k_1 k_5-k_5 k_1 k_3|2}^2}\right)^2\right)\right)\\\eean

The coefficients given above are not the true coefficients yet (except
for $C_{pentagon}$), since of course we need to use the recursion/reduction
formula to get the complete $\eps$ dependence of the coefficients. However,
this is easy to do by replacing $u^n$ with the corresponding form
factors defined in Section 3.3 with the parameters $A,B,C,D$ given
above.

At that point, the non-symmetric expression given above
will also become symmetric (the pentagon coefficient is already symmetric, as it
should be). For example, the $u^2$ term coefficient in
$C_{1|2|34|5}$ is given by $-{ s_{12}^3 s_{51}\over 16
\vev{1~2}\vev{2~3}\vev{3~4}\vev{4~5}\vev{5~1}}$ while in
$C_{1|2|3|45}$ it is given by $-{s_{12}^3 s_{23}\over 16
\vev{1~2}\vev{2~3}\vev{3~4}\vev{4~5}\vev{5~1}}$. After using the appropriate
form
factor, the true coefficient for the box $(1|2|34|5)$ may be expressed as $-\a(\eps){
s_{12}^3 s_{51}^3\over 16
s_{25}^2\vev{1~2}\vev{2~3}\vev{3~4}\vev{4~5}\vev{5~1}}$, and for the box
$(1|2|3|45)$ as $-\a(\eps){s_{12}^3 s_{23}^3\over 16 s_{13}^2
\vev{1~2}\vev{2~3}\vev{3~4}\vev{4~5}\vev{5~1}}$. The latter is
related to the former by index shifting $i\to i+1$, as it must be.

\subsubsection{Confirmation of the result}

Now we compare our result against \cite{Bern:1996ja,Brandhuber:2005jw}, where the basis is
dimensionally shifted. From our result we see
immediately that the part of the amplitude that is reconstructed from
the cut $C_{12}$ is
\bean & &{(K_{12}^2)^2\over 8
\vev{1~2}\vev{2~3}\vev{3~4}\vev{4~5}\vev{5~1}}\gb{2|k_3 k_1 k_5-k_5
k_1 k_3|2} {s_{12}\over 4} { 4^3\over s_{12}^3}I_5[\mu^6] \\
& & -{1\over  \vev{1~2}\vev{2~3}\vev{3~4}\vev{4~5}\vev{5~1}}( s_{51}
s_{12} I_{1|2|34|5}[\mu^4]+ s_{12} s_{23} I_{1|2|3|45}[\mu^4]+
s_{34} s_{45} I_{12|3|4|5}[\mu^4],
\eean
where we have used $u={4\mu^2\over s_{12}}$ and 
the dimensionally shifted basis.
  To compare with equation (15) of
\cite{Bern:1996ja} (or equation (4.1) of \cite{Brandhuber:2005jw})
we need to use  $I_4[\mu^4]=-\eps(1-\eps) I_4^{8-2\eps}$,
$I_5[\mu^6]= -\eps(1-\eps) (2-\eps) I_5^{10-2\eps}$ as well as
\bea tr[\gamma_5\not k_1\not k_2\not k_3\not k_4 ] & = & \gb{2|k_3
k_4 k_1-k_1 k_4 k_3|2}=\gb{2|k_3 k_1 k_5-k_5 k_1 k_3|2}.
\eea
We see that our result agrees exactly with the equation (15) of
\cite{Bern:1996ja}.\footnote{There is  a relative minus sign for the
$I_5^{10-2\eps}$ term because our definition of master integrals does not include the $(-1)^{n}$  used in \cite{Bern:1996ja}.}

\subsection{Four gluons}

In this part, we give only final results (as $u$-integrals)
for four-gluon amplitudes, since the method has already been
elaborated
in the previous five-gluon example.  In principle one then applies our recursion and reduction formulas of Section 3 to find the coefficients.  Here, we choose
instead to confirm our results against those in the literature, which are also given in terms of the final $\mu$-integral, so we do not write the coefficients
explicitly.

To begin with, we must establish our basis.
 For details, see Appendix B.  First, for the zero-mass box, we
 have for example with the  cut $K_{12}$
that $K=K_{12}$, $P_1=K_1$ and $P_2=-K_4$, thus
\bea A & = &  s_{13} s_{41}/4,~~~B=D^2,~~~~C=-s_{41}/2-
s_{12}/4,~~~D=-s_{41}/2, \eea
and so
\bea C[I_{4;12}^{(0m)}] & = & -{2\over s_{41} s_{12}} \int_0^1
du~ u^{-1-\eps}{1\over \sqrt{1+\W A u}} \ln \left( {\sqrt{1+\W A
u}+\sqrt{1-u}\over \sqrt{1+\W A u} -\sqrt{1-u}}\right), ~~~~~~\W A=
{s_{13} \over  s_{23}}.~~\label{0m-cut-gen}
\eea
Second, there are only one-mass triangles. For $(12|3|4)$ with the cut
$K_{12}$ we have the expression
\bea C[I_3(K_{12}, K_4)]= - {1\over s_{12}}\int_0^1 du~ u^{-1-\eps}
\ln \left( {1+\sqrt{1-u}\over 1-\sqrt{1-u}}\right).
\eea
Bubbles are simply $\sqrt{1-u}$ in all cases. We will
compare our results with known results given first by
\cite{Bern:1995db}, in the form given in \cite{Brandhuber:2005jw}.

\begin{itemize}

\item (1) For the helicity configuration $(++++)$ and cut $C_{12}$ we find
\bea C_{12} & = & {s_{12}^2[1~2][3~4] \over 8
\braket{1~2}\braket{3~4}}
\left(-{2\over s_{41} s_{12}} \int_0^1
du~ u^{-1-\eps} u^2{1\over \sqrt{1+\W A u}} \ln \left( {\sqrt{1+\W A
u}+\sqrt{1-u}\over \sqrt{1+\W A u} -\sqrt{1-u}}\right)\right),~~~\label{4-posi}\eea
The integral in parentheses, with its additional factor of $u^2$, is
related to the box integral $K_4$ of \cite{Brandhuber:2005jw}. Using
$u={4\mu^2\over s_{12}}$, we get immediately $ {2[1~2][3~4] \over
\braket{1~2}\braket{3~4}} K_4$.

\item (2) For $(-+++)$ with cut $C_{41}$ we have
\bea & & I_{cut}  = { [2~3]^2 [4~3]^2 \over 4 s_{12} [1~3]^2}
\int_0^1 du~ u^{-1-\eps} \left( -{ 2 s_{13} ( s_{13}- s_{12}
)\over s_{41}^2} u\sqrt{1-u} \right. \nonumber \\
& & \left.+ {2 s_{12}( s_{41}-s_{13})\over s_{41}^2}u \ln \left( {
1+\sqrt{1-u}\over 1-\sqrt{1-u}}\right) + {( 2+ \W A u)u\over
\sqrt{1+ \W A u}}\ln \left( { \sqrt{1+\W A u}+\sqrt{1-u}\over \sqrt{1+\W A
u}-\sqrt{1-u}}\right)\right),~~\W A={s_{13} \over s_{12}}.
~~~\label{I-total}
\eea
To compare with the  results in the literature, we must change
coordinates  via $u={4\mu^2\over s_{41}}$.  We end up with
\bean & & -{2 [2~3]^2 [3~4]^2 \over   [1~3]^2} { s_{13} ( s_{13}-
s_{12} )\over s_{12} s_{41}^3} J_2(s_{41})-{ [2~3]^2 [3~4]^2 \over
[1~3]^2}  \left(  J_4 +{2 s_{13}\over s_{41} s_{12}} K_4\right) + {2
[2~3]^2 [3~4]^2 \over  [1~3]^2} { ( s_{41}-s_{13})\over s_{41}^2}
J_3(s_{41}).
\eean
We find complete agreement with equation (3.17) of
\cite{Brandhuber:2005jw}.\footnote{ 
To confirm agreement of these formulas, one must remember the relative minus
sign for the triangle integral in the basis of \cite{Brandhuber:2005jw}.}

\item (3) For $(--++)$, the cut $C_{12}$ is almost the same as for $(++++)$, just
  multiplied by a factor of ${\vev{1~2}^2 \over [1~2]^2}$.  This is
  enough to get the correct box coefficient.  For the
  cut $C_{41}$ we get
\bea I &=&  { \vev{1~2}^2[3~4]^2 \over s_{41}s_{12}}
\int_0^1 du~ u^{-1-\eps} \nonumber  \\ & & \left[ {u^2
(1+\W A)^2 \over 4\sqrt{1+\W Au}} \ln \left( { \sqrt{1+\W A
u}+\sqrt{1-u} \over \sqrt{1+\W A u}-\sqrt{1-u}}\right) + {\sqrt{1-u}
\over 6} 
(2-5u-3 \W A u)
 \right], ~~~~\W A={s_{13}\over
s_{12}}.~~~\label{I4adj} \eea
It is straightforward to check this result against
\cite{Brandhuber:2005jw}:  the term with the logarithm translates to
$K_4$, the simple bubble is $I_2$, and the terms in the brackets
with $\sqrt{1-u}(u)$ translate to $J_2(s_{41})$.  Again we confirm
agreement.\footnote{We need to use the following dimensional shift identities for the
  basis of \cite{Brandhuber:2005jw}:
\bean I_4^{6-2\eps} & =& -2 J_4-{s t \over 2 u} I_4 - {s\over u}I_3(s)- {t\over u} I_3(t),
\\
I_3^{6-2\eps}(s) & = & {1\over 2} I_2(s)-J_3(s), \\
 I_2^{6-2\eps}(s)& = &  -{2\over 3}
J_2(s)+{s\over 6} I_2(s)\eean }

\item (4) For $(-+-+)$ with cut $C_{41}$ we have
\bea I & = &  { 2 \vev{1~3}^2 s_{13}\over \vev{2~4}^2 s_{41}
}
\int_0^1 du~ u^{-1-\eps}\left[ {\sqrt{1-u} (12+ 3 \W A(6+u) +\W A^2 (4+5 u)) \over 12 \W
A^2} \right. \nonumber  \\ & & + { (1+\W A)^2 ( 8+ 8 \W A u+ \W A^2
u^2)\over 8 \W A^3 \sqrt{1+ \W A u}} \ln \left( { \sqrt{1+\W A
u}+\sqrt{1-u} \over \sqrt{1+\W A u}-\sqrt{1-u}}\right) \nonumber \\
& & \left. -{ (1+ \W A)^2 (2+ \W A u)\over 2 \W A^3 }\ln \left(
{1+\sqrt{1-u}\over 1-\sqrt{1-u}}\right)\right],~~~~~~~\W
A={s_{13}\over s_{12}}.~~~\label{I4alter}\eea
This integral agrees with equation  (3.69) of \cite{Brandhuber:2005jw}
after accounting
for differences of convention.

\end{itemize}
%

\acknowledgments

We would like to thank F. Cachazo for collaboration in the early
stages of this project, and the participants of the HP${}^2$ workshop
for their reception.  
RB and BF are grateful to ETH Zurich for
repeated hospitality. RB acknowledges the stimulating environment of
the Simons Workshop 2006. CA is supported by the Swiss National Fund
under contract NF-Projekt 20-105493. RB is supported by Stichting
FOM. BF and PM are supported by the Marie-Curie Research Training
Network under contracts MRTN-CT-2004-005104 and MEIF-CT-2006-024178.

\appendix

\section{Kinematics}

In this paper we analyze unitarity cuts in Minkowski space with
signature $(+,-,-,-)$.  The kinematic region in question is the one where
$K^2 >0$ and all other invariants are negative.
Let us study the consequences of these conditions in terms
of the four-dimensional momenta $\W \ell_1$ and $\W \ell_2$ of the cut propagators.
These vectors satisfy
\bea \W \ell_1^2=\W \ell_2^2=\mu^2,~~~~~~K-\W \ell_1=\W
\ell_2.~~~\label{Dyn-cond-1}\eea
First, we can choose a frame such that $\vec{K}=(K,0,0,0)$ and ${\W \ell_1}=
(a, b,0,0)$.  Then, ${\W \ell_2}= (K-a, -b,0,0)$.
 The mass-shell
conditions become $a^2-b^2=\mu^2= (K-a)^2-b^2$, so $a={K/ 2}$
and $b^2={K^2/ 4}-\mu^2$. Since $b$ is real, $b^2 \geq 0$, so we draw
the following important conclusion:
\bea  \mu^2\leq {K^2\over 4},~~~\label{Dyn-res-1}\eea
or equivalently,
\bea
u \leq 1.~~~\label{range-u}
\eea
In the procedure described in this paper, we decompose $\W \ell_1 =\ell_1+ z K$ with
$\ell_1^2=0$. Under this decomposition, we can write $\W \ell_1=(b+ z
K,\a b, 0,0)$ with $\a=\pm 1$. Using $b^2={K^2/ 4}-\mu^2$,
$a=b+z K= {K/ 2}$, we can get $z={(1\pm \sqrt{1-u})/ 2}$.
Furthermore, since
we choose the positive light cone with $\delta^+(\ell^2)$, i.e., $b>0$, we have our second
important conclusion: if $K>0$ we need to choose $z={(1-
\sqrt{1-u})/ 2}$, but if $K<0$ we need to choose $z={(1+
\sqrt{1-u})/ 2}$. Throughout the paper, we will always assume $K>0$,
thus
\bea
z={1- \sqrt{1-u}\over 2}.
\eea
 The choice of this solution does not affect our
discussion.

\section{Special cases of master integrals}

\subsection{One-mass and two-mass triangles}
Consider a cut triangle in the massless limit
 where $K_3^2=0$ (so it is a one-mass or two-mass triangle).
From (\ref{I3m-para}), we see that
$Z=1$ and $\sqrt{\Delta_3}=- (2K_1\cdot K_3)= K_1^2- K_2^2$. Thus we
have
\bea C[I_{3}^{1m/2m}(K_1)]=-\int_0^1 du~ u^{-1-\eps}{1\over
\sqrt{\Delta_3}}\ln \left( {1 +\sqrt{1-u}\over 1-\sqrt{1-u}
}\right).
~~~\label{I-3-massless-1}\eea
We can integrate by parts to get a different expression:
\bea \sqrt{\Delta_3}C[I_{3,cut}^{1m/2m}] & = & { u^{-\eps}\over
\eps}\ln \left.\left( {1 +\sqrt{1-u}\over 1-\sqrt{1-u} }\right)\right|_0^1+
\int_0^1 du~{ u^{-\eps}\over \eps} { 1\over u\sqrt{1-u}} \nonumber \\
& = & {1\over \eps}\int_0^1 du~u^{-1-\eps} {1\over
\sqrt{1-u}}.
~~\label{I3m-massless} \eea
Comparing this formula with (\ref{I-2m-another}), we see that the
form is the same. In fact, if we allow  coefficients of scalar functions to
be  general functions of $\eps$, then
there is no need to distinguish one-mass and two-mass triangles from bubbles.
Thus, if one
likes, one can think in terms of keeping only bubble functions in the basis and discarding
both one-mass triangles and two-mass triangles.\footnote{ More concretely, we know that scalar bubbles, one-mass triangles
and two-mass triangles all have the form $c(\eps) (-K^2)^{-\eps}$
where $c(\eps)$ is a function of $\eps$. This same ``modified basis'' has been used in \cite{Britto:2005ha}.}

It is, of course, easy to carry out the $u$ integral in (\ref{I3m-massless}) explicitly and check it against the known expressions for one- and two-mass triangles after restoring the correct normalization factors.

\subsection{Some boxes with massless legs}

Here we discuss some special cases of boxes with massless legs.
We follow all the conventions of Section 3.  Suppose that $P_1$ is the momentum of a massless leg, so $P_1^2=0$.
Then, with the definitions (\ref{Para-3}), we find
Thus for the special case where $P_1^2=0$,   we have $B=
D^2$, and thus
\bean
D - C u
+ \sqrt{1-u}\sqrt{B - A u}
& = & {D\over 2}\left(2-u{ 2C \over D}  + {\rm
  sign}(D)\sqrt{1-u}\sqrt{  1-u{A\over D^2}}\right).
\eean
The expression in parentheses is a complete square, if
\bean 1+ {A\over D^2} = {2C \over D},
\eean
or, equivalently,
\bean
 D^2+A- 2 C D=0.\eean
With the definitions  (\ref{Para-3}) subject to $P_1^2=0$,
\bea
D^2+A- 2 C D & = & { (P_1\cdot K)^2 P_2^2 (K-P_2)^2\over
(K^2)^2}. ~~\label{Para-rela-4}\eea
We see that this expression vanishes
 if $P_2^2=0$
or $(K-P_2)^2=0$. Under this condition, the cut (\ref{B-cut-gen})
takes the following special form.
\bea C[I_{4}(K; P_1, P_2)] & = & {1\over K^2 D} \int_0^1 du~
u^{-1-\eps} {1\over \sqrt{1-u{A\over D^2} }} \ln \left( {
\sqrt{1-u{A\over D^2}}+\sqrt{1-u}\over \sqrt{1-u{A\over
D^2}}-\sqrt{1-u}}\right).~~~\label{B-cut-spe-1}\eea
Here we needed the conditions
\bea P_1^2=P_2^2=0,~~~~{\rm {\bf or}}~~~P_1^2=(K-P_2)^2=0.~~\label{Cond-spe-1}
\eea

\vspace{0.3cm}
{\bf Zero-mass box function:}

For this case we have all $K_i^2=0$, so there are only two cuts, $K_{12}$
and $K_{23}$.  These are trivially related  by index permutation.
For cut $K_{12}$ we have $K=K_{12}$, $P_1=K_1$ and $P_2=-K_4$.
%
Define
\bea
~~~\a=
{K_{13}^2 \over  K_{23}^2}.
\eea
We can see that
\bean B - A u & = &  \left( -{K_{41}^2\over 2}\right)^2 (1+ \a u) \\
D - C u & = & \left( -{K_{41}^2\over 2}\right) \left(1- u(1+{
K_{12}^2\over 2 K_{41}^2}) \right), \eean
thus
\bean  D - C u \pm \sqrt{1-u} \sqrt{B - A u} & = & -{K_{41}^2\over
4} \left( \sqrt{1+ \a u}\pm \sqrt{1-u}\right)^2.\eean
Using this we have
\bea C[I_{4,0m}(K_{12};K_1,-K_4)] & = & -{2\over K_{41}^2 K_{12}^2}
\int_0^1 du~u^{-1-\eps}{1\over \sqrt{1+\a u}} \ln \left( {\sqrt{1+\a
u}+\sqrt{1-u}\over \sqrt{1+\a u} -\sqrt{1-u}}\right).~~\label{0m-cut-gen-app}\eea
This is exactly the expression that we find in the four-gluon examples
(\ref{0m-cut-gen}).

\vspace{0.3cm}
{\bf One-mass box function:}

We assume that $K_1^2\neq 0$, so there are three cuts, $K_1$, $K_{34}$ and
$K_{23}$. We will neglect details and give only results.

For cut $K_1$ we have
\bea C[I_{4,1m}(K_1; K_{12},-K_4)] & = & {2\over K_{34}^2 K_{23}^2}
\int_0^1 du~u^{-1-\eps}{1\over \sqrt{1+\a u}} \ln \left( {\sqrt{1+\a
u}+\sqrt{1-u}\over \sqrt{1+\a u} -\sqrt{1-u}}\right),~~\a={K_1^2
K_{24}^2\over  K_{23}^2 K_{34}^2}.
~~\label{1m-cut-1}\eea

For cut $K_{34}$ we have
\bea C[I_{4,1m}(K_{34}; K_3,-K_2)] & = & -{2\over K_{34}^2 K_{23}^2}
\int_0^1 du~u^{-1-\eps}{1\over \sqrt{1+\a u}} \ln \left( {\sqrt{1+\a
u}+\sqrt{1-u}\over \sqrt{1+\a u} -\sqrt{1-u}}\right),~\a=
{ K_{24}^2 \over K_{23}^2.}~~\label{1m-cut-2}\eea

For cut $K_{41}$ we have
\bea C[I_{4,1m}(K_{41}; K_4, -K_3)] & = & -{2\over
K_{34}^2 K_{23}^2} \int_0^1 du~u^{-1-\eps}{1\over \sqrt{1+\a u}} \ln \left(
{\sqrt{1+\a u}+\sqrt{1-u}\over \sqrt{1+\a u} -\sqrt{1-u}}\right),
~\a= {K_{24}^2 \over K_{34}^2}.~~\label{1m-cut-3}\eea

\vspace{0.3cm}
{\bf Two mass easy box functions:}

We assume $K_1^2\neq 0$ and $K_3^2\neq 0$.  Then there are four possible
cuts.  For each one, it is possible to choose $P_1,P_2$ such that
the condition (\ref{Cond-spe-1}) is satisfied, as shown in the following table.

\bea
\begin{array}{|c|c|c|} \hline
Box~Cut~K & ~~~P_1~~~ & ~~~P_2~~~ \\
\hline K_1 & -K_4 & K_{12} \\
\hline K_3 & -K_2 & K_{34} \\
\hline K_{12} & -K_4 & K_1 \\
\hline K_{23} & K_2 & -K_1 \\
\hline
\end{array}~~~\label{2me-table}
\eea
%

\subsection{Zero-mass pentagon}

Here we evaluate (\ref{Pentagon-gen}) for the zero-mass pentagon under the cut
$K_{12}$. It is
\bea  & & C[I_{5,0m} (K_{12}; K_1,-K_{45}, -K_5)]  = \int_0^1 du
u^{-1-\eps} {1 \over K_{12}^2} \times~~~\label{I5-0m-cut} \\ & & \left(
-{ 4 \gb{2|k_3 k_4 k_5 + k_5 k_4 k_3|2}\over 4 K_{51}^2 K_{23}^2
K_{34}^2 K_{45}^2- \gb{2|k_3 k_1 k_5-k_5 k_1 k_3|2}^2 u} {1\over
\sqrt{1+ \W A  u}}
\ln {\sqrt{1+\W A  u}-\sqrt{1-u}\over \sqrt{1+\W A  u}+\sqrt{1-u}}\right. \nonumber \\
& & + { 4 \gb{3|k_4 k_5 k_1+ k_1 k_5 k_4|3}\over 4 K_{51}^2 K_{23}^2
K_{34}^2 K_{45}^2- \gb{2|k_3 k_1 k_5-k_5 k_1 k_3|2}^2  u} {1\over
\sqrt{1+\W B  u}} \ln {\sqrt{1+\W B  u}-\sqrt{1-u}\over \sqrt{1+\W B
u}+\sqrt{1-u}} \nonumber \\
& & \left. + { 4\gb{5|k_1 k_2 k_3+ k_3 k_2 k_1|5}\over 4 K_{51}^2
K_{23}^2 K_{34}^2 K_{45}^2- \gb{2|k_3 k_1 k_5-k_5 k_1 k_3|2}^2 u}
{1\over \sqrt{1+ \W C  u}}\ln {\sqrt{1+ \W C  u}-\sqrt{1-u}\over
\sqrt{1+ \W C  u}+\sqrt{1-u}}\right), \nonumber \eea
where
\bea \W A = {K_{52}^2\over K_{51}^2},~~~ \W B ={K_{13}^2\over
K_{23}^2},~~~ \W C ={K_{12}^2 K_{35}^2\over K_{34}^2 K_{45}^2}. \eea
%

\subsection{Hexagons and beyond are not independent}

 The integrand of double cut of hexagon is of the form
\bean {\delta(\W\ell^2-\mu^2) \delta( (\W\ell-K)^2-\mu^2) \over
((\W\ell-P_1)^2-\mu^2)
((\W\ell-P_2)^2-\mu^2)((\W\ell-P_3)^2-\mu^2)((\W\ell-P_4)^2-\mu^2)
}.
\eean
The momentum vectors $K,P_i$ as well as $\W
\ell$ are four-dimensional, and moreover the four $P_i$
are linearly independent in general.
Therefore we can express $K$ as a linear combination of the $P_i$:
\bea K= \sum_i \a_i P_i.~~~~~\label{K-exp}\eea
Within the integral we may make the following substitutions:
\bean & & \sum_i \a_i ((\W\ell-P_i)^2-\mu^2) =  \sum_i \a_i (P_i^2
-2P_i \cdot \W \ell) 
 =  \sum_i \a_i P_i^2- 2K \cdot \W \ell  = \sum_i \a_i
P_i^2-K^2.
\eean
 In the first step we used the delta
function $\delta(\W \ell^2-\mu^2)$. In the second step we used
(\ref{K-exp}) while in the third step we have used the second delta
function $\delta( (\W \ell-K)^2-\mu^2)=\delta( K^2-2K\cdot \W
\ell)$.

Using this result we can write
\bean & &\left( {\sum_i \a_i P_i^2-K^2\over \sum_i \a_i
P_i^2-K^2}\right){\delta(\W\ell^2-\mu^2) \delta( (\W\ell-K)^2-\mu^2)
\over ((\W\ell-P_1)^2-\mu^2)
((\W\ell-P_2)^2-\mu^2)((\W\ell-P_3)^2-\mu^2)((\W\ell-P_4)^2-\mu^2)
}\\ & = & {1\over \sum_i \a_i P_i^2-K^2}{\delta(\W\ell^2-\mu^2)
\delta( (\W\ell-K)^2-\mu^2)\sum_i \a_i ((\W\ell-P_i)^2-\mu^2) \over
((\W\ell-P_1)^2-\mu^2)
((\W\ell-P_2)^2-\mu^2)((\W\ell-P_3)^2-\mu^2)((\W\ell-P_4)^2-\mu^2)
}\\ & = & {1\over \sum_i \a_i P_i^2-K^2}\sum_i \a_i
{\delta(\W\ell^2-\mu^2) \delta( (\W\ell-K)^2-\mu^2) \over
\prod_{j\neq i} ((\W\ell-P_j)^2-\mu^2) }.
\eean
Now each term is seen to be a cut pentagon.  The lesson is
that there are no further independent cuts of scalar functions beyond pentagons.

\section{Factors of the form $\vev{\ell| QP |\ell}$}

In spinor manipulation, we repeatedly encounter factors like
$\vev{\ell|Q_iK|\ell}$ and $\vev{\ell|Q_iQ_j|\ell}$.  It is worth
developing a systematic approach to deal with these factors.  Let us
consider a general factor of this type, written as $\vev{\ell| QP |\ell}$.

{\bf Method One: Spinor Basis}

One way to find the poles from this factor is by expansion in a
basis of any two independent spinors:

\bea
\ket{\ell}= \ket{a}+y \ket{b}. \eea
Then the roots of the equation $0=\vev{\ell| QP
|\ell}$
lie at the solutions to the quadratic equation
 \bean 0 & = & \vev{a|QP|a}+ y( \vev{a|QP|b}+\vev{b|QP|a}) +y^2 \vev{b|QP|b}, \eean
which are
\bea
 y_\pm & = & {-( \vev{a|QP|b}+\vev{b|QP|a})\pm \vev{a~b}\sqrt{\Delta}\over 2 \vev{b|QP|b}},
 \eea
 where
 \bean \Delta & = & 4 [ ( Q\cdot P)^2- Q^2 P^2].
 \eean

With these two solutions $\ket{\ell_+},\ket{\ell_-}$ we have
\bea \vev{\ell|QP|\ell}= \vev{\ell~\ell_+}\vev{\ell~\ell_-}
{\vev{b|QP|b}\over \vev{a~b}^2}.
\eea

{\bf Method Two: Vector Solutions}

Here we describe a second approach, which avoids having to choose basis spinors and helps manipulate a variety of expressions.

 Given two massive
momenta $Q,P$ we can construct two lightlike momenta $P_+, P_-$
by solving
\bea (Q+ x P)^2=0,~~\Longrightarrow x_{\pm}={-2 Q \cdot P\pm
\sqrt{4 ( (Q\cdot P)^2 -Q^2 P^2)} \over 2 P^2}= {-2 Q
\cdot P\pm \sqrt{\Delta} \over 2 P^2}.~~~\label{K1K3-x}\eea
We have the following relations among these variables:
\bea
P_{\pm} & = & Q+ x_{\pm} P,~~~\label{P1P2}\\
P &= & {P_+-P_-\over (x_+ -x_-)},~~~~~Q={-x_- P_+ + x_+ P_-\over
(x_+ - x_-)}  \\  x_+ x_- & = & {Q^2 \over P^2},~~~~x_+ +x_-=-{2 Q\cdot
P \over P^2},~~~~x_+ - x_- =x_+ - x_-={\sqrt{\Delta} \over P^2},
\eea
\bea \gb{P_+|Q|P_+} & = & {x_+ \over (x_+-x_-)} (-2 P_+\cdot
P_-)= {x_+ \over (x_+-x_-)} { \Delta\over P^2},\\
\gb{P_+|P|P_+} & = & -{1 \over (x_+-x_-)} (-2 P_+\cdot
P_-)= -{1 \over (x_+-x_-)} { \Delta\over P^2},\\
\gb{P_-|Q|P_-} & = & -{x_- \over (x_+-x_-)} (-2 P_+\cdot
P_-)= -{x_- \over (x_+-x_-)} { \Delta\over P^2},\\
\gb{P_-|P|P_-} & = & {1 \over (x_+-x_-)} (-2 P_+\cdot P_-)= {1
\over (x_+-x_-)} { \Delta\over P^2}.~~~\label{Inner-P1P2}\eea
%

\subsection{Application of second method}

It is easy to check that
\bea \vev{\ell|Q P |\ell} & = &  {1\over (x_+-x_-)}
\vev{\ell~P_+}[P_+~P_-]\vev{\ell~P_-}.~~~\label{K1K3-exp-P}\eea
This means, in particular, that $P_+, P_-$ are exactly the two poles within the factor $ \vev{\ell|Q P |\ell}$.

When we try to identify the structure of the logarithmic part, we
often encounter the following two combinations.
\bea \left( {\gb{P_+|Q|P_+}\over \gb{P_+|P|P_+}}\right)\left(
{\gb{P_-|Q|P_-}\over \gb{P_-|P|P_-}}\right) & = & x_+ x_-= {Q^2\over P^2},
~~~\label{Multiply} \\
\left( {\gb{P_+|Q|P_+}\over \gb{P_+|P|P_+}}\right)\left(
{\gb{P_-|Q|P_-}\over \gb{P_-|P|P_-}}\right)^{-1} & = & {x_+\over
x_-} = { Q \cdot P- \sqrt{  (Q\cdot P)^2 -Q^2
P^2}\over  Q \cdot P+ \sqrt{  (Q\cdot P)^2 -Q^2
P^2}}.~~~\label{Division}\eea
Of these two arguments of logarithms, the one given in (\ref{Multiply}) is unphysical
and so must drop
 out of the final result,
while the one given in (\ref{Division}) is the physical singularity
identifying a given
triangle, box or pentagon.

Sometimes we need to use the spinor components of
$P_+, P_-$. For this we can expand in a basis of two arbitrary spinors,
\bean
\la_{P}= {\ket{a}+ w \ket{b}\over \sqrt{t}},~~~~~~~\W
\la_{P}={|a]+ \O w |b]\over \sqrt{t}},
\eean
 where $t$ is a
normalization factor. We can then solve to find
\bean t={\gb{b|a|b}\over \gb{b|P|b}},~~~~ w= -{ \gb{a|P|b}\over
\gb{b|P|b}}\Longrightarrow \la_{P}= -{ P|b] \vev{a~b}\over
\gb{b|P|b} \sqrt{t}}. \eean

We must also consider factors such as
\bea  \vev{P_+ |R P |P_+} & = &\vev{P_+ |R { P_+-P_-\over
x_+-x_-} |P_+} = -{ \vev{P_-~P_+}\over x_+-x_-} \gb{P_+|R|P_-}, \\
\vev{P_- |R P |P_-} & = &\vev{P_- |R { P_+-P_-\over x_+-x_-} |P_-}
= -{ \vev{P_-~P_+}\over x_+-x_-} \gb{P_-|R|P_+}.\eea
Now we can do the following sum, which is the pattern we
encounter in cut pentagons.
\bean S[Q, P, S,R]& \equiv & {\vev{P_+|R P|P_+}\over
\vev{P_+|S P|P_+}}+
 {\vev{P_-|R P|P_-}\over \vev{P_-|S P|P_-}} =
 { \gb{P_+|R|P_-}\over \gb{P_+|S|P_-}}+  { \gb{P_-|R|P_+}\over \gb{P_-|S|P_+}}\\
 & = & { \gb{P_+|R|P_-}\gb{P_-|S|P_+}+ \gb{P_+|S|P_-}\gb{P_-|R|P_+}\over
 \gb{P_+|S|P_-}\gb{P_-|S|P_+}} \\
 & = & { (2P_+\cdot R) (2P_-\cdot S)+(2P_-\cdot R) (2P_+\cdot S)-(2P_+\cdot P_-)
 (2 R\cdot S) \over (2P_+\cdot S) (2P_-\cdot S)-S^2  (2P_+\cdot P_-)}.
\eean
If we expand $P_+, P_-$ in terms of $Q, P$, then we find that this
quantity can be expressed as follows:
\bea  & & S[Q, P, S,R]  =  {T_1\over T_2},~~~\label{Func-S} \eea
with
\bea
T_1 = -8 \det \left( \begin{array}{lcr} R \cdot S & Q \cdot R & P \cdot R\\
Q \cdot S & Q^2 & Q \cdot P \\ P \cdot S & Q \cdot P &
P^2
\end{array} \right); ~~~~
T_2 = -4 \det \left( \begin{array}{lcr} S^2 & Q \cdot S & P \cdot S\\
Q \cdot S & Q^2 & Q \cdot P \\ P \cdot S & Q \cdot P &
P^2
\end{array} \right).~~~~\label{T1-T2}
\eea
The function $T_1$ is symmetric under exchange of the first two, or the last two, arguments of $S$.  The function $T_2$ depends only on $Q,P,S$ and is symmetric in all three.

\subsection{Spinor integral formulas}

Here we derive some useful spinor integral formulas.

First let us consider
\bean \int \vev{\ell~d\ell}[\ell~d\ell]{1\over \gb{\ell|K|\ell}
\gb{\ell|Q |\ell}} = \int_0^1 dx \int
\vev{\ell~d\ell}[d\ell~\partial_\ell]\left({ [\eta~\ell] \over
\gb{\ell|R|\ell} \gb{\ell|R|\eta}}\right),~~~~R=x Q+(1-x)K.\eean
Now we take $\eta$ to be one solution of $\vev{\ell|QK|\ell}$, i.e,
the solution (\ref{K1K3-x}) in which we have substituted $P\to K$, so
\bean R= x{ -x_- P_++x_+ P_-\over x_+-x_-}+(1-x) { P_+-P_-\over
x_+-x_-}, \eean
where $x_{\pm}$ are defined as in (\ref{K1K3-x}) and $P_{\pm}$ as in  (\ref{P1P2}). Taking $\eta$ to be $P_+$
we have
\bean \gb{\ell|R|\eta}= \gb{\ell|R|P_+} = \vev{\ell~P_-}[P_-~P_+]
\left( x { (x_++1) \over (x_+-x_-)} -{1\over (x_+-x_-)} \right).\eean
Now using
\bea \int_0^1 dx {1\over (x c+d) (x a+b)}= -{1\over a d-bc}\left( \ln
{c+d\over d}-\ln{a+b\over b }\right),~~~\label{x-pattern-0-first}\eea
we  get
\bea
\int_0^1 dx \left({ [\eta~\ell] \over \gb{\ell|R|\ell}
\gb{\ell|R|\eta}}\right) &=& {(x_+-x_-)[P_+~\ell]\over
\vev{\ell~P_-}[P_-~P_+] \gb{\ell|P_+|\ell}}\ln \left(
{-x_+\gb{\ell|K|\ell}\over \gb{\ell|Q|\ell}}\right) \nonumber \\
&=& -{1\over
\vev{\ell|Q K|\ell}}\ln \left( {-x_+\gb{\ell|K|\ell}\over
\gb{\ell|Q|\ell}}\right),~~\label{rep-rul-1}\eea
where we have used the formula (\ref{K1K3-exp-P}) to simplify the
result.\footnote{It is important to realize that in principle we
should also take the pole contribution of $\vev{\ell~P_-}$ from the
middle equation of (\ref{rep-rul-1}). However, in many examples,
there is a factor of $\vev{\ell|Q K|\ell}$ in the numerator, so this pole
has zero residue.}


\end{document}